\documentclass{article}
\PassOptionsToPackage{table}{xcolor}

\usepackage{arxiv}
\usepackage{CJKutf8}
\usepackage[utf8]{inputenc} 
\usepackage[T1]{fontenc}    
\usepackage{hyperref}       
\usepackage{url}            
\usepackage{booktabs}       
\usepackage{amsfonts}       
\usepackage{amsmath}
\usepackage{nicefrac}       
\usepackage{microtype}      
\usepackage{lipsum}
\usepackage{graphicx}
\usepackage{xcolor}
\usepackage{natbib}
\usepackage{multirow}
\usepackage{array}
\usepackage[table]{xcolor}
\usepackage{pifont}
\usepackage[most,breakable]{tcolorbox} 
\graphicspath{ {./images/} }
\usepackage{color, soul}

\title{MagicGUI: A Foundational Mobile GUI Agent with Scalable Data Pipeline and Reinforcement Fine-tuning}

\author{
  \renewcommand{\arraystretch}{1.8} 
  \begin{tabular}{c@{\hspace{.65cm}}c@{\hspace{.65cm}}c@{\hspace{.65cm}}c@{\hspace{.65cm}}c}
    Liujian Tang \textsuperscript{1} \thanks{These authors contributed equally to this work.} &
    Shaokang Dong \textsuperscript{1} \footnotemark[1] &
    Yijia Huang \textsuperscript{1}  &
    Minqi Xiang \textsuperscript{1}  &
    Hongtao Ruan  \textsuperscript{1}  \\
    Bin Wang \textsuperscript{1}  &
    Shuo Li \textsuperscript{2} &
    Zhiheng Xi \textsuperscript{2} &
    Zhihui Cao \textsuperscript{1} &
    Hailiang Pang \textsuperscript{1}  \\
    Heng Kong \textsuperscript{1} &
    He Yang \textsuperscript{1}  &
    Mingxu Chai \textsuperscript{2} &
    Zhilin Gao \textsuperscript{1} &
    Xingyu Liu \textsuperscript{1}  \\
    Yingnan Fu \textsuperscript{1} &
    Jiaming Liu \textsuperscript{1} &
    Xuanjing Huang \textsuperscript{2} &
    Yu-Gang Jiang \textsuperscript{2} &
    Tao Gui \textsuperscript{2} \footnotemark[2]  \\
    \multicolumn{5}{c}{Qi Zhang \textsuperscript{2} \footnotemark[2] \quad \quad \quad
    Kang Wang \textsuperscript{1} \footnotemark[2] 
    \quad \quad \quad
    Yunke Zhang \textsuperscript{1} \footnotemark[2] 
    \quad \quad \quad
    Yuran Wang \textsuperscript{1} \thanks{Corresponding authors: Tao Gui (tgui@fudan.edu.cn), Qi Zhang (qz@fudan.edu.cn), Kang Wang (wangkang12@honor.com), Yunke Zhang (zhangyunke@honor.com), and Yuran Wang (wangyuran1@honor.com).} }
  \end{tabular} \\
  \\
  \textit{\textsuperscript{1} Honor Device Co., Ltd} \vspace{.2cm} \\
  \textit{\textsuperscript{2} Fudan University}
}

\begin{document}
\maketitle
\begin{abstract}
    \renewcommand{\thefootnote}{\fnsymbol{footnote}}
This paper presents MagicGUI, a foundational mobile GUI agent designed to address critical challenges in perception, grounding, and reasoning within real-world mobile GUI environments. The framework is underpinned by following six key components: (1) \textbf{a comprehensive and accurate dataset}, constructed via the scalable GUI Data Pipeline, which aggregates the largest and most diverse GUI-centric multimodal data to date from open-source repositories, automated crawling, and targeted manual annotation; (2) \textbf{enhanced perception and grounding capabilities}, facilitating fine-grained multimodal alignment for UI element referencing, grounding, and screen comprehension; (3) \textbf{a comprehensive and unified action space}, encompassing both fundamental UI operations and complex interactive intents to support human-agent interactions; (4) \textbf{planning-oriented reasoning mechanisms} that enable the model to decompose complex user instructions into sequential actions with explicit intermediate meta-paln reasoning; (5) \textbf{an iterative two-stage training procedure}, combining large-scale continue pre-training on 7.8M samples with reinforcement fine-tuning utilizing a spatially enhanced composite reward and dual filtering strategy; and (6) \textbf{competitive performance} on both the proprietary Magic-RICH benchmark and over a dozen public benchmarks, achieving superior performance across GUI perception and agent tasks, while demonstrating robust generalization and real-world deployment potential in practical mobile GUI scenarios, as detailed in Figure~\ref{fig:1}.\footnotemark[4]\footnotetext[4]{Our proprietary Magic-RICH benchmark and the MagicGUI model are available in \url{https://huggingface.co/GUIAgent}. In addition, we release the inference and evaluation scripts at \url{https://github.com/MagicAgent-GUI/MagicGUI/tree/main}.}
\renewcommand{\thefootnote}{\arabic{footnote}}
\end{abstract}

\section{Introduction}

\begin{figure}[!t]
  \centering
  \includegraphics[scale=0.3]{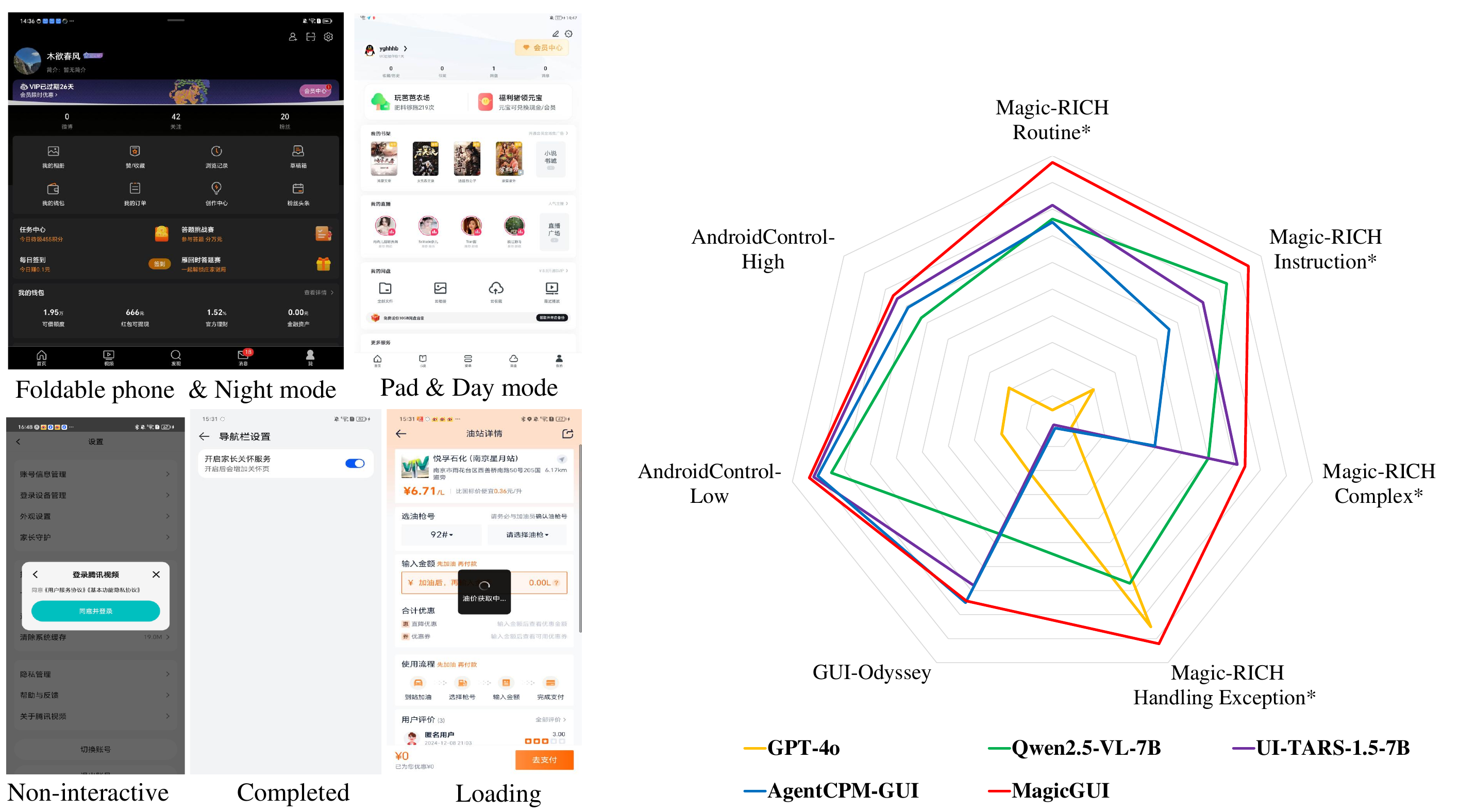}
  \caption{(a) Representative samples of our proposed Magic-RICH benchmark. (b) Performance comparison.}
  \label{fig:1}
  \vspace{-2mm}
\end{figure}

With the rapid advancement of Multimodal Large Language Model (MLLM), these models have demonstrated remarkable improvements in perceptual and reasoning capabilities ~\citep{schneider2024foundation,Qwen2-VL,chen2024internvl,Cogvlm}. They are now able to perceive visual information from the real world more accurately encompassing text, images, and videos, while also emulating human-environment interaction behaviors with greater fidelity. The enhancement of these fundamental capabilities has spurred increasing research on applying MLLMs to tackle a wide range of real-world tasks. A prominent research is enabling MLLM-based agents to seamlessly interact with graphical user interfaces (GUIs) for automated execution of user commands, thereby ushering in a new era of GUI auto-agents~\citep{wang2024survey,nguyen2024gui}.

The GUI agent is conceived as a computational entity capable of interpreting complex human instructions and, through perception and reasoning, automating task execution in digital environments with minimal human intervention by leveraging graphical elements (e.g., icons, buttons, input fields) and common interface tools. Previous GUI agents consistently employ a modular hybrid approach, achieving rapid development of domain-specific tasks through the integration of text representation modules for structured extraction from GUI interfaces (e.g., web HTML, app XML), understanding and reasoning modules (e.g., GPT-4o~\citep{hurst2024gpt}) and memory storage modules~\citep{wang2024oscar,wu2024copilot,zhang2024ufo,xie2024osworld}. However, this approach is inherently dependent on expert knowledge, specialized VLMs, and constrained operational environments. Even minor variations in tasks, vision-language models (VLMs), or runtime environments can induce systemic failures~\citep{xia2024agentless,Aguvis} in the overall framework, demonstrating significantly less scalability and adaptability compared to contemporary mainstream end-to-end agent architectures.

The end-to-end agent framework~\citep{Os-atlas,cogagent,UI-TARS,AgentCPM-GUI,gui_r1,wu2025gui} integrates the comprehension of user instructions, perception of GUI interfaces, task reasoning, and execution within a unified model. The model demonstrates human-like adaptive capabilities across diverse tasks and operational contexts, exhibiting remarkable scalability. It reveals substantial potential to evolve into a universal artificial intelligence assistant for all digital ecosystems.

\begin{figure}[h!]
  \centering
  \includegraphics[scale=0.46]{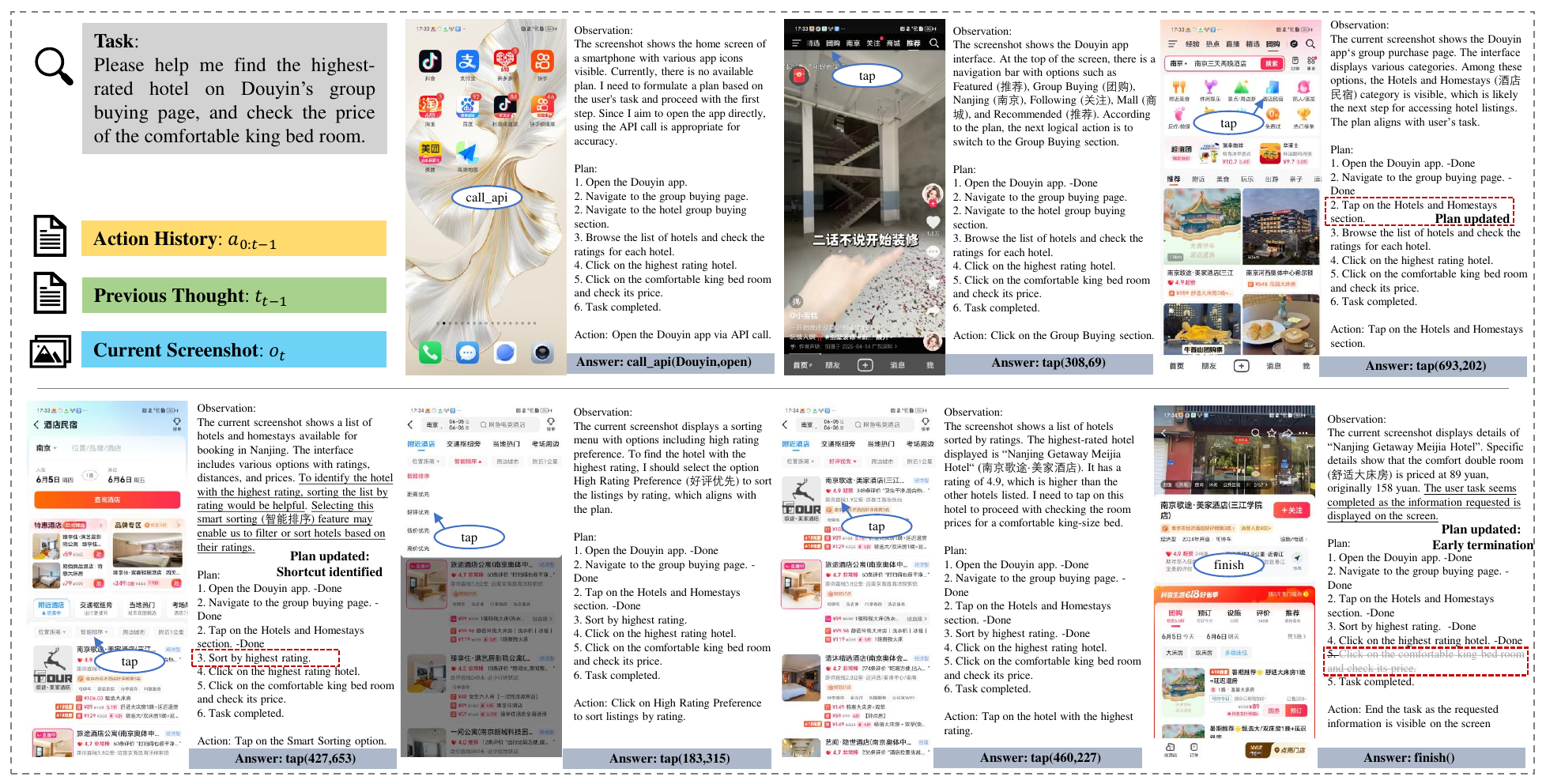}
  \caption{A demo case study of MagicGUI to check the price of the comfortable king bed room.}
  \label{fig:case_gui}
  \vspace{-2mm}
\end{figure}

Despite significant advancements in GUI agent development, current auto-agents still face several critical challenges: (1) \textbf{Data Scale and Quality}: Existing open-source datasets are limited in application coverage and suffer from inherent noise issues. Collecting large-scale, high-quality and multi-language user trajectory data remains highly challenging, while automated simulation-based data collection or synthetic data generation also struggles with inevitable noise. (2) \textbf{Perception Optimization}: GUI environments exhibit substantial heterogeneity in UI styling, page layout, and information density. This heterogeneity poses significant difficulties for agents to maintain fine-grained perceptual accuracy across all UI interfaces, especially when UI elements are extremely small, numerous, and densely packed. (3) \textbf{Reasoning Generalization}: Agents are required to demonstrate generalized reasoning and execution capabilities across various GUI environments, including the ability to formulate adaptive operation sequences based on environmental characteristics and dynamically adjust action strategies in response to contextual variations.

To address these challenges, we propose a foundational mobile GUI agent, \emph{\textbf{MagicGUI}}, characterized by robust generalization and adaptive reasoning capabilities, as demonstrated in several cases in Figure ~\ref{fig:case_gui}. Specifically, MagicGUI presents the following contributions:

\begin{itemize}
\item \textbf{Sufficient and Accurate Dataset}: We propose a scalable and modular GUI Data Pipeline to collect high-quality dataset for mobile GUI perception and grounding across our HONOR N, HONOR X, HONOR Magic, and HONOR Flip series devices. We leverage a large volume of open-source general data, such as Infinity~\citep{gu2024infinity} and ScreenQA~\citep{hsiao2024screenqa}, for general capabilities. Integrated with diverse open-source mobile GUI datasets, including OS-ATLAS~\citep{Os-atlas}, AMEX~\citep{amex}, \mbox{GUIAct~\citep{guiact}}, AndroidControl~\citep{ac} and GUI-Odyssey~\citep{odyssey}, this dataset establish a foundational milestone to ensure the superior accuracy and generalization performance of our model in both HONOR mobile devices and the open-source leaderboard.

\item \textbf{Perception and Grounding Capabilities}: To enhance the perception and grounding capabilities of the GUI agent model across diverse and complex app layouts, we curate five core types of training data, each addressing a critical aspect of GUI understanding and interaction: (1) Element Referring, which establishes a rigorous classification system for UI elements, enabling precise identification of element types; (2) Element Grounding, focusing on accurate localization of UI elements essential for interaction tasks; (3) Element Description, augmenting the agent’s holistic understanding by integrating multiple feature dimensions into comprehensive element descriptions; (4) Screen Caption, enabling the agent to generate coherent descriptions of entire GUI screens; (5) Screen VQA, enhancing interactive understanding by responding to on-screen questions.

\item \textbf{Comprehensive and Unified Action Space}: To emulate human interactions on mobile devices, we have designed a comprehensive and unified action space applicable across different mobile platforms. In addition to fundamental operations such as Tap, Scroll, Text Input, Navigation Back, Navigation Home, Long Press, Finish, we have incorporated more complex interactive actions, including Wait, Enter, Takeover, Drag, Call API, Screenshot, Long Screenshot, and NoAnswer actions. This comprehensive and unified action space enhances the instruction-following and execution capabilities for high-level user tasks, significantly boosting the model's applicability on mobile devices.

\item \textbf{Planning-Oriented Reasoning}: For high-level agent tasks, we integrate a planning-oriented reasoning mechanism at each inference step. Specifically, the model observes the environment, refines the meta-plan, and selects the subsequent action. To maintain efficiency and prevent excessively long context, the environment is restricted to the current GUI context, historical action traces, and the preceding plan. Consequently, our model achieves improved task-level consistency and more accurate decision-making in dynamic GUI environments.

\item \textbf{Iterative Two-Stage Training Procedure}: We design a two-stage training procedure for the MagicGUI model to balance accuracy on the proprietary Magic-RICH dataset with generalization on open-source datasets. Specifically, (1) Continue Pre-training (CPT) enhances the core perception, grounding, and navigation capabilities; (2) Reinforcement Fine-tuning (RFT) improves robustness and generalization across diverse datasets by introducing a spatially enhanced composite reward function within a unified action space. Additionally, we develop both static and dynamic data filtering methods to eliminate irrelevant or low-value samples.

\item \textbf{Competitive Performance}: We evaluate the referring, grounding, and GUI agent capabilities of the MagicGUI model across widely adopted open-source benchmarks. Additionally, we introduce a proprietary Magic-RICH dataset supporting the Chinese language and native applications. MagicGUI consistently achieves superior performance across all benchmarks, demonstrating robust generalization across diverse scenarios.

\end{itemize}

\section{Related Work}
Recent GUI agent research follows three main paradigms: (1) Prompt-Driven GUI Agent Models (e.g., AppAgent) relying on commercial models, (2) Supervised Fine-Tuning-Based GUI Agent Models (e.g., CogAgent) specializing in grounding tasks and GUI agent tasks, and (3) Reinforcement Fine-Tuning-Based GUI Agent Models (e.g., DigiRL) optimizing action policies. While works like OS-Atlas and UI-TARS advance reasoning, they remain limited by narrow action spaces and static data biases.

\subsection{Prompt-Driven GUI Agent Models}
The advancement of high-performance Multimodal Large Language Models (MLLMs), such as QwenVL~\citep{Qwen2-VL, qwen2.5vl} and the InternVL series~\citep{chen2024far, chen2024internvl}, has paved new pathways for addressing Graphical User Interface (GUI) tasks. Recently, MLLM-based GUI agent frameworks have attracted significant attention. Researchers have enhanced GUI agent capabilities in device control, task execution, and interface comprehension by integrating visual perception with linguistic understanding. AppAgent~\citep{appagent} and Mobile-Agent~\citep{mobileagent} utilize general-purpose commercial models like GPT for GUI task planning and prediction. However, these agents remain heavily reliant on prompt engineering to handle complex tasks. Moreover, they struggle to effectively address challenging and specialized GUI tasks.

\subsection{Supervised Fine-Tuning-Based GUI Agent Models}
For a more domain-specific GUI agent model, several studies have developed end-to-end models to address grounding~\citep{seeclick_screenspot,cogagent,UGround} and GUI agent tasks~\citep{Aria-UI,Showui,Os-atlas,UI-TARS}, integrating fundamental perception, reasoning, and execution capabilities. These approaches are predominantly data-driven and trained via Supervised Fine-Tuning (SFT) on general Visual Question Answering (VQA), grounding, or instruction-level GUI datasets. Specifically, SeeClick~\citep{seeclick_screenspot} introduces a GUI grounding pre-training strategy and subsequently establishes the first realistic GUI grounding benchmark \emph{ScreenSpot}. Moreover, CogAgent~\citep{cogagent} enhances the original CogVLM model~\citep{Cogvlm} by incorporating a novel high-resolution cross-module to manage varying image resolutions. However, these models are confined to grounding tasks and are not applicable to GUI agent operations on real mobile devices.

Furthermore, UGround~\citep{UGround} and Aria-UI~\citep{Aria-UI} introduce an instruction planner that generates specific actions using GPT-4, followed by training a visual grounding model capable of accurately mapping diverse referring expressions of GUI elements to coordinates to execute GUI agent tasks. Besides that, several studies integrate action training with the grounding task. For instance, OS-Atlas~\citep{Os-atlas} operates in three distinct modes, including grounding mode, action mode and agent mode, to accommodate various task requirements. UI-TARS~\citep{UI-TARS} incorporates deliberate system-2 reasoning into multi-step decision making and synthesizes both positive and negative reasoning examples to train SFT and DPO models. However, these methods heavily rely on SFT training procedure with limited datasets, thereby constraining their generalization and applicability across diverse evaluation benchmarks and real mobile devices.

\subsection{Reinforcement Fine-Tuning-Based GUI Agent Models}
Rule-based reinforcement fine-tuning, as previously validated by OpenAI o1~\citep{openai_o1} and DeepSeek-R1~\citep{deepseek_r1}, has demonstrated remarkable generalization performance in the domains of natural logic puzzles~\citep{logic_rl}, mathematical reasoning~\citep{deepseek_math,othink_mr1}, and code generation~\citep{stepcoder}. Subsequently, VLM-R1~\citep{vlmr1} and Visual-RFT~\citep{visual_rft} have extended this training framework to encompass general vision-language tasks, including referring expression comprehension, object detection, and reasoning grounding, by designing a rule-based Intersection over Union (IoU) reward function.

For mobile GUI tasks, DigiRL~\citep{digirl} and DistRL~\citep{distrl} collect online trajectories in mobile simulations and employ another large Vision-Language Model (VLM) to assess whether the current task is completed as a reward signal. Consequently, these models necessitate increased interaction and inference time during the training process, resulting in reduced efficiency. With the emergence of more substantial static GUI datasets~\citep{seeclick_screenspot,screenspot_pro,amex,guiact,ac,odyssey}, showcasing the training procedure on such static datasets and further deployment on real mobile devices has become mainstream. 

For instance, ReachAgent~\citep{reachagent} decomposes high-level tasks into page reaching and operation subtasks. However, it requires the input of XML documents and screenshots of GUI pages, which complicates its widespread application on real devices. Other native GUI agent models, UI-R1~\citep{ui_r1} and GUI-G1~\citep{gui_g1}, focus solely on the low-level GUI grounding task, while GUI-R1~\citep{gui_r1} and InfiGUI-R1~\citep{infigui-r1} are capable of addressing both complex low-level and high-level GUI agent tasks. Nevertheless, these methods directly apply the GRPO/RLOO reinforcement learning algorithms on the Qwen2.5-VL model~\citep{qwen2.5vl} without a cold-start training process, inevitably leading to suboptimal or less accurate performance in specific GUI tasks. Furthermore, these native GUI-agent models concentrate on static datasets with a limited action space and are unable to represent more complex user interactive operations in mobile devices.

\section{Methodology}

\begin{figure}[htbp]
  \centering
  \includegraphics[scale=0.29]{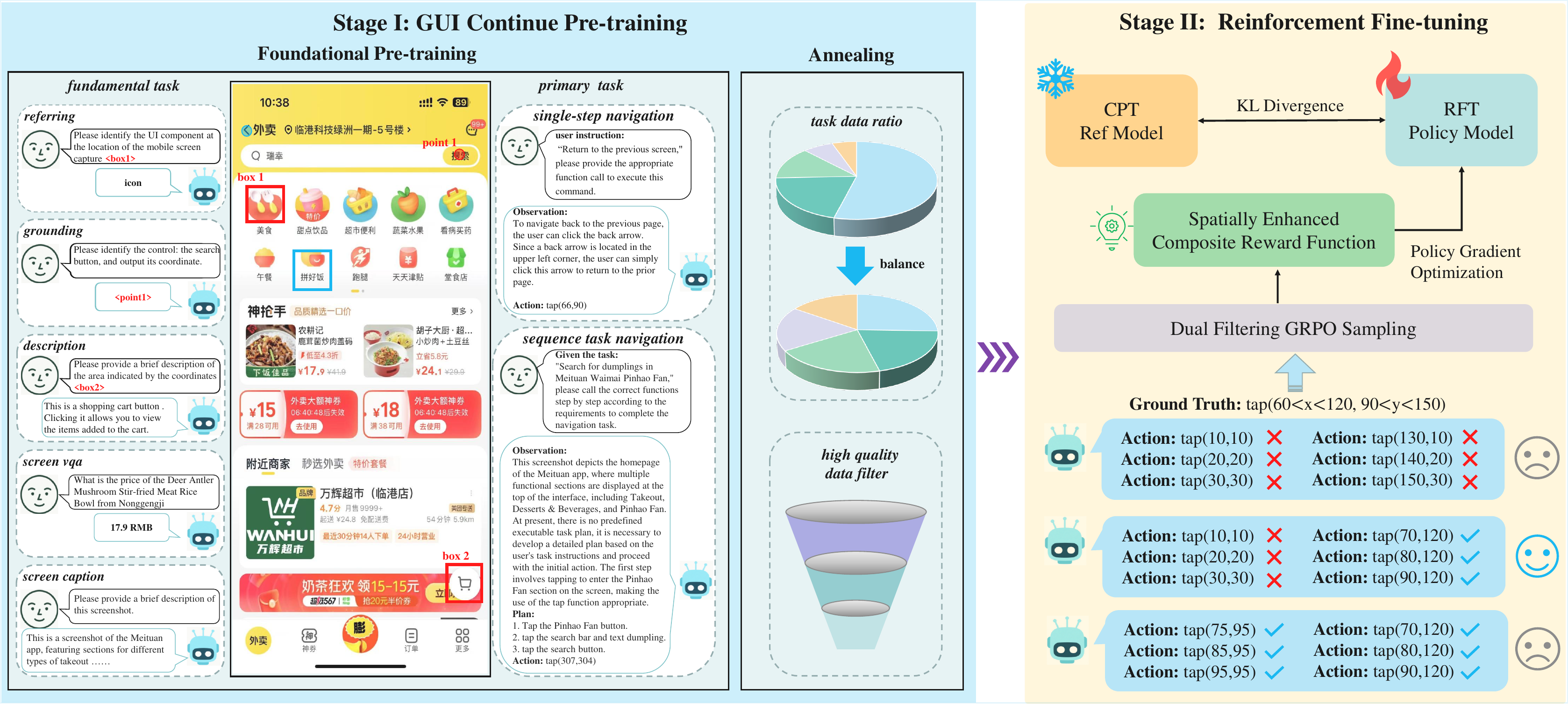}
  \caption{Overview of the MagicGUI framework.}
  \label{fig:magic gui framework}
\end{figure}

In this section, we introduce \emph{\textbf{MagicGUI}}, an end-to-end GUI agent model designed to manage both low-level and high-level GUI tasks autonomously. MagicGUI receives the current GUI screenshot and user query as inputs, generating valid actions and grounding positions when necessary. The overall training framework is depicted in Figure~\ref{fig:magic gui framework}, which is divided into two stages: \textbf{Stage I: Continue Pre-training (CPT)} (detailed in Section \ref{First Stage}), which involves training a foundational model on a large and diverse dataset followed by an annealing phase using a balanced and high-quality dataset; and \textbf{Stage II: Reinforcement Fine-tuning (RFT)} (detailed in Section \ref{Second Stage}), aimed at further enhancing the model’s robustness and generalization capabilities.

\subsection{Continue Pre-training}
\label{First Stage}
The continue pre-training stage consists of two stages: Foundational Pre-training on large-scale GUI-centric data to acquire general perception and grounding capabilities, and Annealing Training on high-quality samples to refine alignment and reduce noise. This process equips the model with both broad coverage and task-specific understanding for GUI interaction.

\subsubsection{Data Format}
\tcbset{
  myboxstyle/.style={
    colback=gray!20,     
    colframe=black!70,    
    coltitle=white,       
    fonttitle=\bfseries,  
     fontupper=\itshape,
    boxrule=0.8mm,       
    arc=1mm,               
    boxsep=1mm,             
    left=1mm,              
    right=1mm,              
    top=1mm,               
    bottom=1mm,             
    toptitle=0mm,           
    bottomtitle=0mm,     
    enhanced,                  
  }
}
Following Qwen2-VL, we use <image> as the image placeholder and add $<|vision\_start|>$ and $<|vision\_end|>$ to mark the image feature sequence boundaries. The training data adopts the ChatML format, where $<|im\_start|>$ and $<|im\_end|>$ denote each utterance. Additionally, inspired by Qwen3~\citep{qwen3tech}, we introduce $<think>$ and $</think>$ to control the reasoning process in the output.
\begin{tcolorbox}[
    myboxstyle, 
    title=Data Format of Continue Pre-training
]
$<$|im\_start|$>$user
$<$|vision\_start|$>$ \{\texttt{image}\} <|vision\_end|> \{\texttt{query}\}$<$|im\_end|$>$.

$<$|im\_start|$>$system
$<$think$>$ \{\texttt{think}\} $<$/think$>$ \{\texttt{answer}\}$<$|im\_end|$>$.
\end{tcolorbox}
\subsubsection{Foundational Pre-training}
We adopt Qwen-VL~\citep{qwen2.5vl} series models as our base models and further train them with GUI data containing approximately one billion tokens. We unfreeze the visual encoder and dynamically adjust the image resolution fed into the visual encoder, thereby enhancing the model's perceptual capability for GUI images.
Throughout the foundational pre-training phase, the model has been fully pre-trained on both general and GUI-related knowledge, with linear warm-up and decay schedules utilized to maintain training stability. 

For the low-level navigation (step-level action) task, our model predicts the optimal action based on a given step-level instruction and the observation of the device screen. This process can be formatted as:
\begin{align}
    action = VLM(instruction,observation).
\end{align}  
For high-level GUI tasks, our model is employed to iteratively observe the screenshots and generate the corresponding actions at each time step throughout the task sequence. This sequential process can be described as
\begin{align}
    a_n=VLM(task,a_{0:n-1},o_{0:n-1},o_n).
\end{align}
where $task$ represents the high-level instruction for the GUI tasks, $a_{0:n-1}$ and $o_{0:n-1}$ denote the sequences of past actions and historical observations (device screenshots) prior to time step $n$, respectively. $o_n$ corresponds to the screen observation at step $n$, and $a_n$ denotes the action that the model is required to predict. To ensure efficient training and inference, we exclusively utilize historical actions as model input, striking a balance between model performance and training cost. Therefore, the optimization objective of our GUI navigation task is reformulated as:
\begin{align}
    a_n=VLM(task,a_{0:n-1},o_n).
\end{align}  
Furthermore, we incorporated planning-oriented reasoning training into foundational pre-training phase based on our findings that the model often struggles to generate correct navigation plans and actions for certain challenging tasks. Deep reasoning can substantially improve performance on these tasks. Moreover, integrating GUI reasoning data during the foundational pre-training phase lays a solid foundation for developing navigation reasoning capabilities in subsequent reinforcement fine-tuning. The optimization objective of the GUI navigation task is reformulated as:
\begin{align}
    t_n,a_n=VLM(task,a_{0:n-1},t_{0:n-1},o_n).
\label{eq:task objective}
\end{align}  
In Equation \eqref{eq:task objective}, $t_{0:n-1}$ denotes the reasoning information produced by the model prior to time step $n$, whereas $t_n$ represents the reasoning text at the current time step $n$.

The loss function formulation for the optimization objective of our model at CPT stage can be formally expressed as:
\begin{align}
    Loss = -\frac{1}{N}\sum_{i=1}^{N}\sum_{j=1}^{V}y_{ij}log(p_{ij}),
\end{align}      
where $N$ is the sequence length, $V$ is the vocabulary size, $y_{ij}$ is the one-hot encoding of the true class at position $i$, and $p_{ij}$ is the predicted probability for class $j$ at position $i$. 

\subsubsection{Annealing}
Annealing on a small amount of high-quality data can improve the performance of the model in key tasks~\citep{li2024datacomp, grattafiori2024llama}. In contrast to the foundational pre-training phase, our annealing employs a limited set of high-quality data and reduces the learning rate by 50\% to facilitate more stable and precise parameter optimization, thereby further enhancing the overall performance of MagicGUI. 

We meticulously sample data by task type from the foundational pre-training phase, selecting a total of 350k high-quality samples by conducting manual sampling and quality inspections. We rebalance the task distribution in the training dataset and increase the proportion of GUI-related data to boost the model's performance on GUI agent tasks, while maintaining its general capabilities. For GUI agent tasks, we additionally apply a data augmentation strategy by upsampling rare UI element and action data to ensure robust performance on long-tail navigation tasks. Evaluation of the annealing impact demonstrates an average improvement of 1.0\% in the foundational pre-training model’s performance.

\subsection{Reinforcement Fine-tuning}
\label{Second Stage}

We model the GUI agent task as a finite-horizon Markov Decision Process (MDP) $M=\{S,A,P,R,\mu_0,H\}$, where $S$ represents the state space comprising screenshots and user queries, $A$ denotes a finite vocabulary token space for the GUI agent to represent valid GUI actions and grounding positions. The state transition function $P: S \times A \times S \rightarrow [0,1]$ defines the probability of transitioning to the subsequent state given the selected action. $R: S \times A \rightarrow \mathbb{R}$ is a rule-based reward function that indicates the validity and accuracy of the GUI action in completing the task. $\mu_0$ represents the distribution of initial states, which varies across different mobile devices or static datasets. $H$ signifies the finite horizon of the episode for each task. For low-level tasks, $H = 1$ and $S$ denotes the state space for screenshots and stepwise instruction, whereas for high-level task, $H > 1$ and $S$ denotes the state space for screenshots and task description. At each time step $t$, the GUI agent observes a state $s_t \in S$, selects an action $a_t \in A$ based on the learned policy $\pi_{\text{RL}}(a_t|s_t;\theta_t)$ and derive the rule-based reward $r_t \in R$. The initial policy for the GUI agent is $\pi_{\text{CPT}}(a_t|s_t)$ through the aforementioned training procedures. Ultimately, the agent's objective is to find an efficient policy to maximize the expected return as $\mathbb{E}_{\pi_{RL}}(\sum_{t=0}^H r_t)$.

\subsubsection{Spatially Enhanced Composite Reward Function in Unified Action Space}
\label{Rule-based Reward}

\begin{figure}[!t]
  \centering
  \includegraphics[scale=0.64]{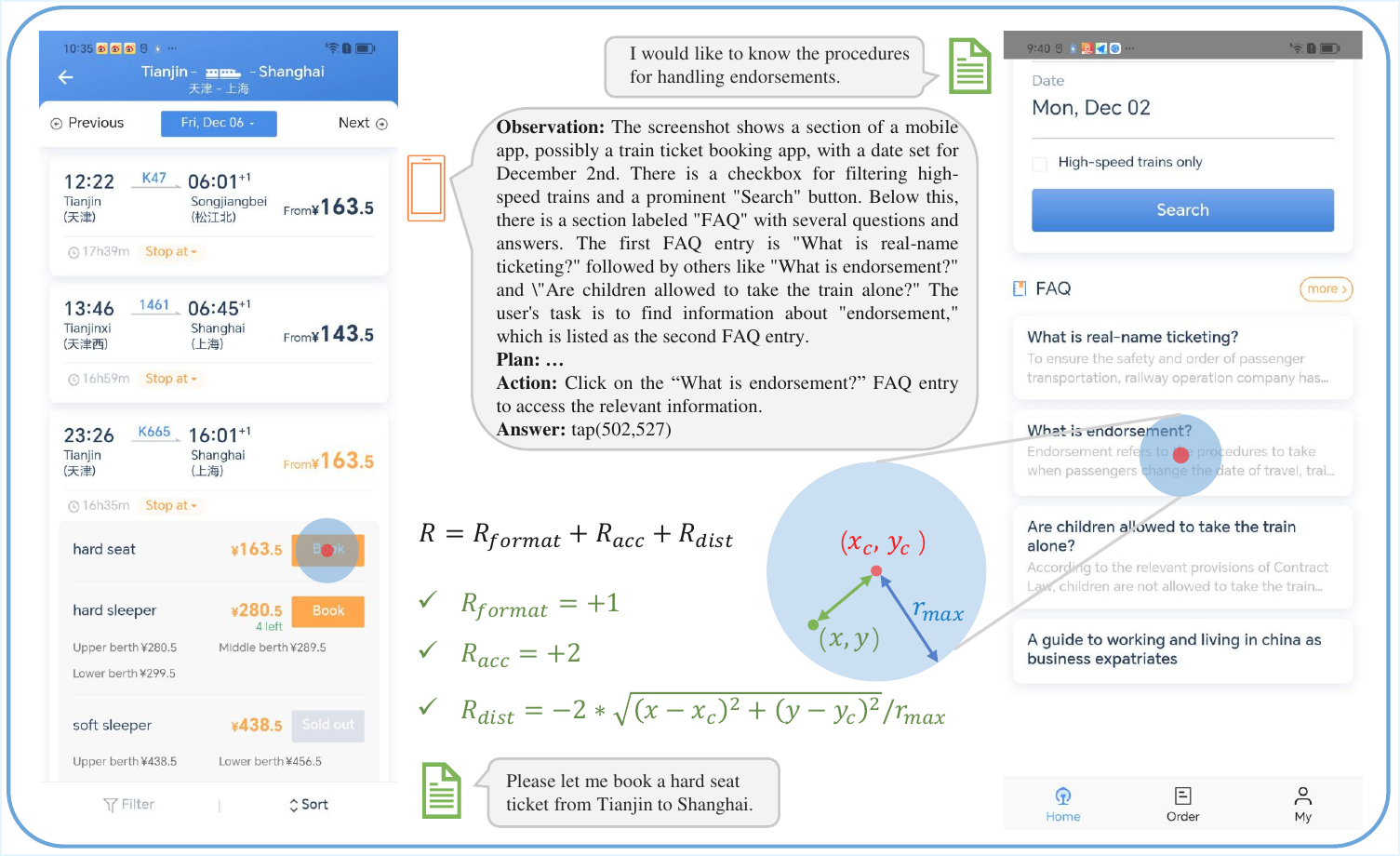}
  \caption{Illustration of the spatially enhanced composite reward function.}
  \label{fig:rm}
\end{figure}

The rule-based reward function has been validated by OpenAI o1~\citep{openai_o1} and DeepSeek-R1~\citep{deepseek_r1}. In our GUI agent tasks, the accurate action type, text input content, and grounding position are three critical aspects for parsing valid executable actions. As shwon in Figure \ref{fig:rm}, our spatially enhanced rule-based reward function is composed of the following three components:
\begin{align}
    R = R_{format} + R_{acc} + R_{dist}.
\end{align}

\textbf{Format reward.} $R_{format} = +1$ if the response format is correct; otherwise, $R_{format} = -1$. For fast reasoning conditions, the format is used to assess the action as $ActionType(coordinate: (x,y), content)$. The $ActionType$ represents the basic action operation, while the coordinate denotes the operation position, and the content includes the text input or scroll direction if necessary. For the condition of planning-oriented reasoning format, the response must generate the required HTML tags as \texttt{<think>}...\texttt{</think>} and \texttt{<answer>} ... \texttt{</answer>}.

\textbf{Accuracy reward.} The accuracy reward $R_{acc} = +2$ if the output action type, coordinates, and content align with the ground truth; otherwise, $R_{acc} = -2$. For the action type, the output must be strictly identical to the ground truth. For coordinates $(x,y)$, since the open-source datasets only contain the ground truth coordinates $(x_c,y_c)$, we define a bounding circle $\mathcal{R}$ centered at $(x_c,y_c)$ with a maximum relative radius $r_{max}$ set to 14\% of the screenshot size. This design is aligned with the evaluation in open-source benchmark~\citep{Os-atlas,UI-TARS}. If the output coordinates $(x,y)$ fall within the bounding circle $\mathcal{R}$, the coordinates are considered aligned with the ground truth. For the drag action, considering the limited valid drag space, we constrain the distance to a maximum of 7.5\% of the screenshot size. Furthermore, for certain operations, including scroll and call api to open/kill apps, the output content must also be strictly identical to the ground truth. In contrast, for text input actions, we compute the F1 score between the predicted text and the ground truth $(gt)$, considering the text correct if $F1 > 0.5$. The details of the accuracy reward for all actions are provided in Table~\ref{tab:acc_reward}, where $dist(\cdot,\cdot)$ function represents the Euclidean distance of two points.

\begin{table}[!ht]
\centering
\caption{Additonal conditions for accuracy reward $\mathbf{R_{acc} = +2}$ in unified action space.}
\label{tab:acc_reward}
\renewcommand{\arraystretch}{1.35}
\begin{tabular}{lp{7.5cm}p{5.2cm}}
\toprule[1.5pt]
\textbf{Action} & \textbf{Description} & \textbf{Conditions for} $\mathbf{R_{acc} = +2}$ \\
\midrule[1pt]
Tap &  Click at coordinate $(x, y)$ & $dist([x,y],[x_c,y_c]) \leq 14\%$ \\
Scroll & Scroll at coordinate $(x, y)$ with $direction$ up / down / left / right & $dist([x,y],[x_c,y_c]) \leq 14\%$ and $direction = gt[direction]$ \\
Text Input & Type $text$ at coordinate $(x, y)$ & $dist([x,y],[x_c,y_c]) \leq 14\%$ and $F1(text, gt[text]) > 0.5$ \\
Navigation Back & Adb command to go back to the previous page & -- \\
Navigation Home & Adb command to go to the home screen of the  mobile  & -- \\
Long Press & Long Press at coordinate $(x, y)$ & $dist([x,y],[x_c,y_c]) \leq 14\%$ \\
Finish & Indicate that navigate task has been completed & -- \\
Wait & wait for several seconds & -- \\
Enter & Adb command to press enter & -- \\
Takeover & Request user takeover & -- \\
Drag & Drag from coordinate $(x_1, y_1)$ to coordinate $(x_2, y_2)$ & $dist([x_1,y_1],[x_{1c},y_{1c}]) \leq 7.5\%$ and $dist([x_2,y_2],[x_{2c},y_{2c}]) \leq 7.5\%$ \\
Call API & Adb command to $open/kill$ app & $app = gt[app]$ and $open/kill = gt[operation]$ \\
Screenshot & Adb command to screenshot & --  \\
Long Screenshot &  Adb command to long screenshot & -- \\
\bottomrule[1.5pt]
\end{tabular}
\end{table}

\textbf{Distance reward.} Test data in open-source datasets often provide ground truth for only a single coordinate and lack actual bounding boxes or bounding circles. Furthermore, the sizes of text and icon components in these datasets differ from those on our Honor mobile devices. Solely applying the accuracy reward based on a bounding circle with an upper-bound relative radius $r_{max}$ may lead to suboptimal performance on open-source mobile GUI datasets. Therefore, we additionally design a distance reward to penalize the deviation between the predicted coordinates $(x,y)$ and the ground truth coordinates $(x_c,y_c)$. In our experiments, we analyze the impact of the distance reward on both proprietary and open-source datasets. The distance reward $R_{dist}$ is defined as follows:
\begin{align}
    R_{dist} =     
    \begin{cases} 
    - 2 * \sqrt{(x-x_c)^2 + (y-y_c)^2} / r_{max}, & \text{if } R_{acc} = +2 \\
    0, & \text{if } R_{acc} = -2
    \end{cases}.
\end{align}

\subsubsection{Algorithm: DF-GRPO}
For GUI agent tasks, we propose the \textbf{Dual Filtering Group Relative Policy Optimization (DF-GRPO)} algorithm, which filters irrelevant or valueless training data following the aforementioned CPT procedure, including \textbf{static filtering} and \textbf{dynamic filtering}. Specifically, given the current state $s$ containing the screenshot and user query, our model generates a group of responses $\{o_i\}_{i=1}^{G}$ for each state $s$. Subsequently, each response is evaluated through the rule-based reward function as $\{r_i\}_{i=1}^{G}$. The overall objective for DF-GRPO to optimize the policy $\pi_{\theta}(a \vert s)$ is defined as follows:
\begin{align}
    \mathcal{J}&_{\text{DF-GRPO}}(\theta) = \mathbb{E}_{s \sim S, \{o_i\}_{i=1}^G\sim \pi_{\theta_\text{old}}(\cdot \vert s)} \nonumber \\
    & \Bigg[\frac{1}{G}\sum_{i=1}^{G}\frac{1}{|o_i|}\sum_{l=1}^{|o_i|} 
    \Bigg(\min \Big( \frac{\pi_{\theta}( o_{i,l} \vert s, o_{i, <l})}{\pi_{\theta_\text{old}}( o_{i,l} \vert s, o_{i, <l})} A_i, \text{clip} \big( \frac{\pi_{\theta}( o_{i,l} \vert s, o_{i, <l})}{\pi_{\theta_\text{old}}( o_{i,l} \vert s, o_{i, <l}) }, 1 - \varepsilon, 1 + \varepsilon \big) A_i \Big) - \beta D_{KL}(\pi_{\theta} \| \pi_{ref})\Bigg) \Bigg]  \nonumber \\
    & \quad\quad \text{s.t.}\quad 0 < \Big|\{r_i \mid r_i > 0 \}\Big|< G \quad \text{and} \quad 0 < \Big|\{r_i \mid r_i < 0 \}\Big|< G,
    \label{DF-GRPO}
\end{align}
where
\begin{align}
    A_{i} = \frac{r_i - {mean}(\{r_1,r_2,...,r_G\})}{{std}(\{r_1,r_2,...,r_G\})}, \quad D_{KL}(\pi_{\theta} \| \pi_{ref}) = \frac{\pi_{ref}( o_i \vert s)}{\pi_{\theta}( o_i \vert s)} - \log \frac{\pi_{ref}( o_i \vert s)}{\pi_{\theta}( o_i \vert s)} - 1.
\end{align}
The mean and standard deviation represent the mean and standard deviation of the group of rewards, respectively. In addition, the KL divergence of the reference CPT policy ensures stable learning in the reinforcement learning process. 

We first implement a \textbf{static filtering} strategy to enhance the quality of training data. This approach removes training samples where the CPT model consistently produces either entirely correct or entirely incorrect predictions across all generated responses. Static filtering eliminates redundant or misleading samples that do not contribute to effective reinforcement learning, allowing the model to focus on more challenging and informative samples.

Following static filtering, we design a \textbf{dynamic filtering} mechanism to ensure more stable and efficient learning in GRPO, as reflected in the constraints $0 < |\{r_i \mid r_i > 0 \}|< G$ and $0 < |\{r_i \mid r_i < 0 \}|< G$ in Eq.(\ref{DF-GRPO}). For instance, in a difficult GUI agent task, the current policy outputs a group of responses, all with the correct format but inaccurate coordinates or content ($r_i = -1, \forall i \in [G]$). The advantage $A_i$ for each response equals zero, while the objective of the original DF-GRPO degrades to the KL penalty of the reference CPT policy. In this condition, the RL policy will be optimized in a meaningless and inefficient direction, leading to unstable training. In the case of an easy task where all rewards $r_i \ge 1$, the rewards differ based on the distance to the ground truth coordinates $(x_c, y_c)$. Optimizing under this condition may lead to overfitting, as there exist some low-quality training data containing incorrect annotations. Together, static and dynamic filtering form a comprehensive dual filtering framework that balances training stability and efficiency, ultimately leading to more robust and generalizable GUI agent policies.

\section{The GUI Data Pipeline: High-Quality Scalable Pipeline for GUI-Centric Multimodal Dataset Construction}

High-quality GUI data construction at scale remains a fundamental bottleneck. Most available datasets (i.e. WidgetCaption~\citep{li2020widget}, GUIAct~\citep{guiact}, AndroidControl~\citep{ac}) suffer from limited diversity and outdated coverage. The scale required for effective multimodal training makes manual annotation infeasible, while preserving both semantic consistency and interaction diversity across large volumes of data remains challenging.

To tackle these challenges, we introduce the GUI Data Pipeline—a scalable and high-quality framework for constructing large-scale GUI-centric multimodal datasets. It consists of four stages: (1) raw data collection, (2) data preprocessing, (3) hierarchical task annotation, and (4) data refinement with automated quality monitoring. Using this pipeline, we build \textbf{the largest and most diverse GUI dataset to date}, spanning a wide range of applications and interaction tasks. 

\subsection{Raw Data Collection}
We collect data from three sources: open-source repositories, automated crawling, and manual collection, covering both UI elements in GUI pages and navigation action sequences. The automated crawling of a large number of applications ensures the diversity of GUI data, while the manual collection of long-tail UI elements and navigation actions helps maintain a balanced GUI dataset.

\subsubsection{Open-source Repositories}

As part of our data collection process, we incorporate data from several existing open-source GUI datasets, including RICO, Screen2Vec, MultiApp, and MoTIF, among others. These datasets offer diverse examples of GUI structures, interaction types, and annotation forms, contributing to the overall diversity of our dataset in terms of screen layout, app domains, and action granularity.

However, despite their value, most open-source datasets suffer from several limitations. Many are outdated, reflecting older interface designs and interaction paradigms. Some datasets exhibit unclear or inconsistent annotation standards, particularly for referring expressions or hierarchical semantics. Others focus narrowly on specific domains (e.g., mobile-only or form-filling tasks), resulting in limited coverage of real-world application scenarios. To overcome these limitations, we further augment and unify the collected data through large-scale crawling.

\subsubsection{Automated Crawling}
\label{actiondatacollection}
In addition to open-source datasets, we perform large-scale automatic crawling to collect diverse and up-to-date GUI data from real-world mobile applications. Our crawling strategy consists of two complementary components.

To mitigate privacy risks, all automated data collection is conducted using dedicated, pre-registered accounts specifically created for this purpose. These accounts do not contain any personal user data, and are isolated from real user environments. This design ensures that no personally identifiable information (PII) or sensitive user content appears in the raw collected data.

\paragraph{Broad-coverage-oriented Data Collection}

We develop an automated collection platform based on Android UI automation tools (e.g., UIAutomator 2), which performs deep traversal of each app’s interface to collect broad-coverage-oriented data. The system systematically explores all reachable pages and records screenshots along with their corresponding XML UI hierarchies, storing them in a structured format. This process enables comprehensive coverage of static GUI states and element layouts across apps.

\begin{figure}[htbp]
  \centering
  \includegraphics[scale=0.5]{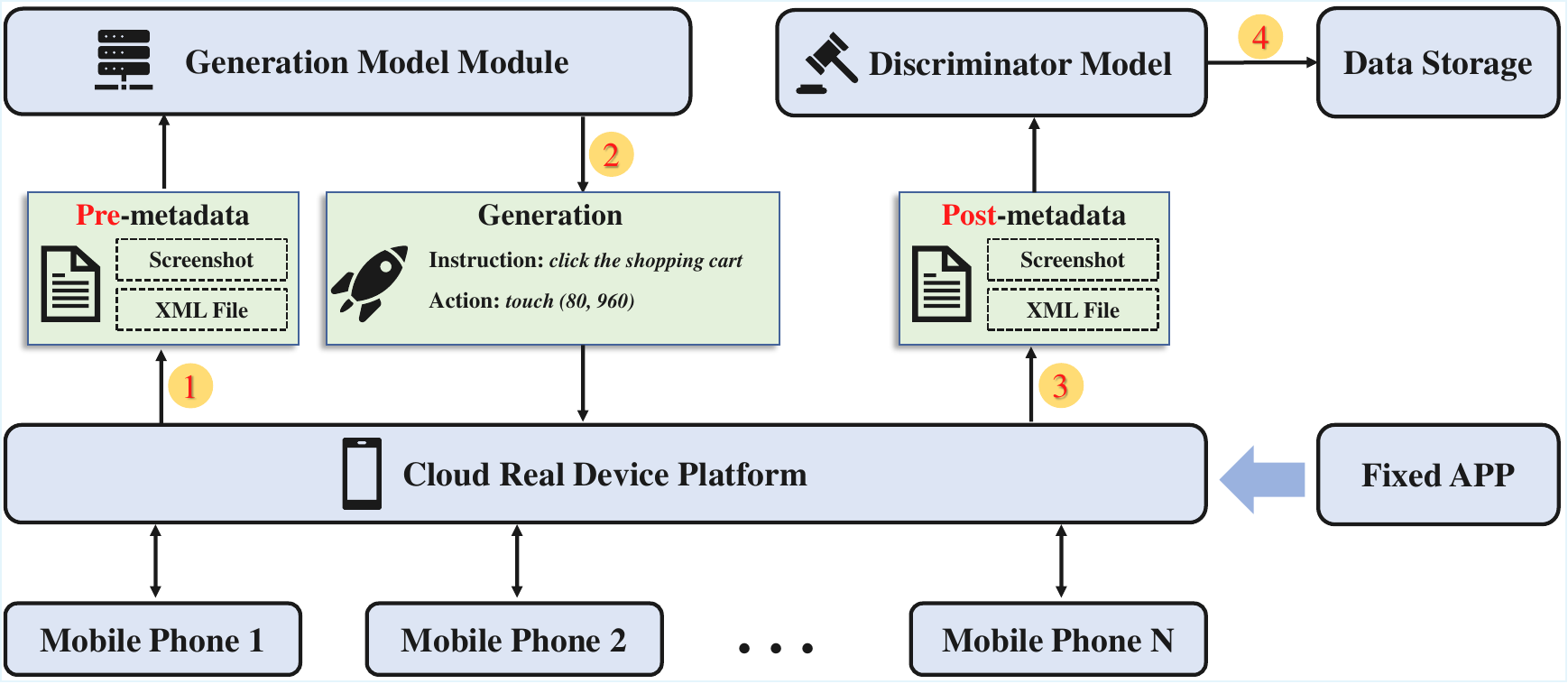}
  \caption{Automatic annotation of single-step action data.}
  \label{fig:automatic_annotation}
\end{figure}

\paragraph{Balanced Element and Action Category Data Collection}
To effectively capture diverse user interactions, we have developed a hardware-software co-optimized framework called the Cloud Real Device Platform, which is designed to collect single-step action data across a wide range of commonly used mobile apps.  As shown in Figure~\ref{fig:automatic_annotation}, the Cloud Real Device Platform connects multiple mobile phones, centrally manages task assignment and status reset of each device, and is responsible for scheduling between action command generation and mobile screen operations. Our data collection workflow consists of four steps as follows:
\begin{itemize}
    \item \textbf{App Pre-metadata Collection.} Real smartphones are connected to the Cloud Real Device Platform, which provides programmable interfaces for both control and data capture. For each fixed app, we first collect the pre-metadata including current screenshot and corresponding XML file.
    
    \item \textbf{Instruction and Action Generation.} Next the app metadata is sent to the Generation Model Module. We utilize a VLM-basd Model Module to generate appropriate action instructions for UI elements based on the current screenshot and XML information. We enhance the sampling ratio of long-tail UI elements and increase the proportion of instructions generated for long-tail action categories to ensure a balanced distribution of action instructions. Then, we use a multimodal large language model to generate the corresponding actions of these instructions.
    
    \item \textbf{App Post-metadata Collection.} The instruction and action are subsequently sent back to the Cloud Real Device Platform. Mobile phones would automatically execute the action. The post-metadata including the screenshot and XML after the execution is collected for the next step.
    
    \item \textbf{Data Storage.} Finally, we use a discriminator to assess the retention of the data samples. The discriminator would estimate if the action achieves the intention of instruction on mobile phones according to the pre-metadata and post-metadata. Each saved sample includes a pre-metadata, a post-metadata, an instruction, and its corresponding action, supporting grounded multimodal learning at the action level.
\end{itemize}

\subsubsection{Manual collection}
\label{Manual collection}
To enhance coverage of rare cases, we manually collect a targeted subset of GUI samples focusing on corner cases that are difficult to capture through automated crawling. These include screens with nested pop-up windows, dynamically rendered elements, semantically ambiguous layouts, and non-standard interaction patterns such as long-press or swipe-only actions. Annotators manually record the screenshot, XML hierarchy, and aligned instructions or action.
\subsection{Data Preprocessing}
Following large-scale data collection, we perform a series of preprocessing steps to improve data quality and consistency. This includes noise data filtering, duplicate data removal, and unified category definition for UI elements and actions, ensuring a clean and standardized dataset for downstream multimodal learning.

\subsubsection{Noise Data Filtering}
To improve data quality before downstream processing, we apply a two-stage noise filtering strategy that combines both rule-based and model-based methods. Together, these methods provide a robust mechanism for reducing label noise and structural errors in the collected data.

\paragraph{Rule-Based Filtering}
We design a set of heuristic rules to eliminate samples with obvious structural or semantic errors. These include GUI pages with missing or corrupted screenshots, incomplete or malformed XML hierarchies, invalid or duplicate element attributes (e.g., missing bounds or undefined class names), and unbalanced or empty element trees. We also filter out screens with unusually sparse or dense structures (e.g., fewer than 2 or more than 100 visible elements), as these often correspond to non-informative states such as splash pages or blank transitions. In addition, we discard referring instructions containing spatial expressions that contradict the actual element layout, or action samples with inconsistent action-type labels.
\paragraph{Model-Based Filtering}
Beyond rule-based heuristics, we leverage open-source MLLMs (e.g., Qwen2.5-VL) and close-source models' APIs to assign semantic quality scores to generated samples. These models assess alignment between instructions, layouts, and actions. Samples with low alignment scores or ambiguous predictions are discarded.

\subsubsection{Duplicate Data Filtering}

To reduce redundancy, we perform duplicate data filtering across both static and dynamic samples. For screenshots, we apply perceptual hash-based comparison to identify visually identical or near-identical images. For XML structures, we compare layout trees using normalized string representations to detect structural duplicates. Additionally, for instruction-action pairs, we compute embedding-based semantic similarity to remove repeated or paraphrased instances. This step effectively eliminates excessive repetition and helps maintain the diversity and informativeness of the dataset.

\subsubsection{Unified Category Definition for UI Element and Action}
In line with existing studies, we apply a unified category definition for UI elements and actions in the training data to maintain semantic consistency across samples.

\paragraph{Unified Element Category}
The absence of standardized UI element categorization in open-source widget datasets primarily stems from dependency on app XML files, which exhibit frequent inconsistencies in element labeling. To address this, we abandon XML-defined classifications and develop a novel taxonomy grounded in visual and functional attributes, including background properties, border characteristics, OCR and interactive behaviors. This methodology enables the systematic definition of twenty UI element types, categorized into two groups, the fundamental elements encompass Text, Icon, Button, Input Field, Switch, Select Box, Page Indicator, Image and Slider, the container elements comprise Option Area, Scroll Picker, Multi-scroll Picker, Date Picker, Navigation Bar, Pop-up Window, Slider Container, Progress Bar, Advertisement, Notification and Map. More details about UI elements are presented in Table \ref{tab:element categories}.

\begin{table}[!ht]
\centering
\caption{UI element categories}
\label{tab:element categories}
\renewcommand{\arraystretch}{1.4}
\begin{tabular}{lp{5cm}@{\hspace{0.5cm}}|@{\hspace{0.5cm}}lp{5cm}}
\toprule[1.5pt]
\textbf{Fundamental} & \textbf{Description} & \textbf{Container} & \textbf{Description}\\
\midrule[1pt]
Text & Display textual information or use for click & Opiton Area & Area containing multiple homogeneous functional UI elements \\
Icon & Abstract graphic or symbol for display or click & Scroll Picker & Scroll to select internal options \\
Button & Combination of text and icon for click & Multi-scroll Picker & Combination of multiple scroll picker \\
Input Field & Area for entering text & Date Picker & Area of selectable days of a month \\
Switch & Click to toggle ON and OFF states & Navigation Bar & Area containing multiple navigation sections, can be scrolled \\
Select Box & Click to toggle selection or deselection & Pop-up Window & Window popping up from and overlaying the main page \\
Page Indicator & Indicate the current page and enable rapid swipe to specified page & Advertisement & Display advertisements or promotional information \\
Image & Larger graphic showing detailed information & Progress Bar & Show progress, most can be dragged \\
Slider& Button on slider track, dragging to adjust corresponding info & Slider Container & Area containing slider and slider track \\
 &  & Notification & System or app notifications popping up and disappearing automatically \\
 &  & Map & Show map, can be dragged and scrolled \\
\bottomrule[1.5pt]
\end{tabular}
\end{table}

\paragraph{Unified Action Space}
Our work systematically addresses the standardization of atomic interaction gestures across diverse mobile platforms, including bar-style phones, flip phones, and tablets. The proposed action space not only incorporates fundamental operations (Tap, Scroll, Text Input, Navigation Back, Navigation Home, Long Press, Finish) but also introduces seven rare operations (Wait, Enter, Takeover, Drag, Call API, Screenshot, Long Screenshot) that remain underrepresented in current open-source datasets, yet critical for comprehensive interaction modeling. The categories of actions and their detailed descriptions are delineated in the aforementioned Table \ref{tab:acc_reward}.

We have decomposed and merged the actions present in open-source datasets to ensure consistency of the action space among open-source data and in-house data. For example, the “search” action in open-source data was decomposed into two distinct actions: Text and Enter. Furthermore. We introduce the Takeover action to facilitate seamless human intervention in authorization and verification scenarios. To address challenges such as network exceptions and countdown advertisements encountered during task execution, we also incorporate the Wait action, enhancing the robustness and reliability of the overall process. 

\subsection{Hierarchical Task Annotation}
As a means to provide structured supervision for grounded GUI understanding, we design a hierarchical annotation framework that mirrors the decision-making process of interactive agents. It consists of three levels: (1) perception and grounding, aligning UI elements with textual references; (2) action prediction, selecting executable operations; and (3) reasoning, modeling the inference steps between instructions and actions. This structure facilitates fine-grained multimodal training and supports more interpretable agent behavior.

\subsubsection{Perception \& Grounding Task Annotation}
Given that effective execution of GUI tasks requires not only precise element-level perception for accurate localization but also comprehensive global scene understanding for holistic interpretation of screenshots, our framework introduces five core tasks specifically designed to:

\textbf{Element Referring.} Different categories of UI elements exhibit different interaction patterns, and open-source data often suffer from issues such as missing element types or conflicting element definitions. To address this, we establish a rigorous and visually grounded definition and classification system for UI element categories. We develop an element classification model (~\citep{lu2024omniparser}) to accurately identify element types, such as Icon, Button, Pop-up Window, Slider, and Scroll Picker. Moreover, PaddleOCR is integrated to perform text recognition within or around these elements. Furthermore, we design an automated module to generate the basic meanings and functionalities of each element. Consequently, a dataset is obtained that details the type, fundamental meaning, and surrounding OCR text of each element, enabling agents to better perceive page layouts and accurately distinguish between various UI elements.

\textbf{Element Grounding.} Localization capability is crucial for executing GUI tasks, particularly actions such as tapping, text input, and dragging. Batch parsing and filtering of XML files are performed to extract extensive element coordinates. Additionally, the dataset is enriched by incorporating long-tail data generated through object detection pre-labeling and manual verification. Subsequently, open-source models (such as Qwen2-VL~\citep{Qwen2-VL} and UI-TARS~\citep{UI-TARS}) are utilized to synthesize the grounding dataset. To assist these models in locating and perceiving accurately, the set-of-marks approach~\citep{yang2023set} is employed to mark the positions of these elements within the images. To ensure the diversity and richness of the generated data, Ferret-UI~\citep{you2024ferret} is referenced, and a variety of prompt templates encompassing multiple perspectives, including absolute positioning, relative positioning, ordinal relationships, functionality, and semantics~\citep{Showui} are meticulously designed.

\textbf{Element Description.} Whether in referring tasks or grounding tasks, each task data typically contains only limited observational information about the UI elements. To enhance the agent’s comprehensive perception of UI elements and better capture intrinsic relationships between different dimensions of the features, multiple dimensions of information are integrated, including element category, color, shape, spatial position, and element functionality, to build a more complete descriptive dataset for UI elements. Specifically, the spatial position encompasses both the absolute and relative locations of elements within the screenshot, while functionality refers to the potential user intentions and the functions that can be achieved through interaction with the screen. A combination of rule-based and model-based approaches is employed to assess multiple dimensions of the generated descriptions, discarding data with low confidence scores. Additionally, the open-source model is required to predict the spatial positions of UI elements based on these descriptions, which are then compared against ground truth to further eliminate low-quality data.

\textbf{Screen Caption.} The objective of constructing the screen captioning task is to enable the agent to achieve a comprehensive understanding of the entire GUI screen by capturing the intricate relationships between the individual elements and the hierarchical structure of the page. This capability serves to mitigate the agent’s hallucinations during screen comprehension, facilitate the differentiation of distinct screens, and identify variations of identical elements across different pages. The idea is drawn from the article “From the Least to the Most”~\citep{cheng2024least}, and a bottom-up approach is employed to construct the page description. Initially, descriptions for all individual elements within the page are generated. Subsequently, these descriptions are progressively integrated layer by layer, leveraging the hierarchical structural information of the page, ultimately yielding a comprehensive and coherent representation of the entire page.

\textbf{Screen VQA.} VQA is a capability that can be directly perceived and interacted with by the user. While the base model possesses VQA capability, a large amount of actions or auxiliary data can diminish the model's understanding of on-screen Q\&A information. To address this challenge, prompt engineering is employed to distill knowledge from open-source models (such as UI-TARS and Qwen2-VL) in the field of visual question answering on mobile terminals through a multi-step process. This data includes unique experiential knowledge and relative positional relationships within each app, as well as relevant negative example data. Specifically, to maintain the model's ability to refuse to answer VQA questions, corresponding unanswerable questions are generated based on the presence of relevant information in the images during this process. All the aforementioned VQA data undergoes repeated answering and semantic evaluation by multiple models to ensure the accuracy and logical consistency of the questions and answers. The original images of the data primarily originate from various popular domestic apps, ensuring a smooth question-answering experience for users on these software platforms.

\subsubsection{Action Task Annotation}

The model is trained on four types of action-related tasks: low-level actions, medium-level actions, high-level actions, and state-transition actions. These tasks are designed with increasing levels of difficulty, enabling the model to progressively acquire a deeper and more comprehensive understanding of user interactions within GUI environments.

\textbf{Low-level Action.} Queries in this task consist of straightforward instructions that directly interact with visible UI elements on the current interface, such as “click the search button” or “drag the music progress bar to the end.” By integrating UI component knowledge acquired from auxiliary tasks like grounding and VQA, the model achieves better consistency and alignment between instruction understanding and action execution. An instruction generation model based on Qwen2-VL-7B~\citep{Qwen2-VL} is further fine-tuned to produce low-level instructions conditioned on the current page and ground-truth actions. Additionally, as described in Section~\ref{Manual collection}, the dataset is enhanced with manually annotated instructions to ensure sufficient coverage of rare actions.

\textbf{Medium-level Action.} Medium-level instructions are more complex than low-level ones, as they often require a deeper understanding of current UI elements. For example, a query such as “I want to check my rewards” may not correspond to a directly visible “Rewards” component on the current page but instead requires navigating through menus like “My” or “Menu.” These tasks further enhance the model’s capability in navigation. To generate such instructions, the model is prompted using chain-of-thought (CoT) reasoning, comparing the pages before and after an action to identify the new information present only on the post-action page. Medium-level instructions are then generated based on these differences. Observations indicate that models \textbf{without specialized training}, when processing both pre- and post-action pages, often struggle with long-context reasoning, which frequently leads to hallucinations. Therefore, high-quality data is synthesized using GPT-4o, and a medium-level instruction generation model is fine-tuned to achieve a good balance between accuracy and efficiency.

\textbf{High-level Action.} Queries in this task represent user-level goals that typically require a sequence of actions to accomplish, as demonstrated in works such as AMEX~\citep{amex}, GUIAct~\citep{guiact}, AndroidControl~\citep{ac}, and GUI-Odyssey~\citep{odyssey}. The MagicGUI model is trained using both manually annotated action traces and curated open-source traces, ensuring that the action space is well-aligned across different datasets.

\textbf{State-transition Action.} To further enhance the model’s understanding of page transitions, a setup similar to UI-TARS~\citep{UI-TARS} and MobileVLM~\citep{mobilevlm} is adopted. Specifically, the model is presented with two consecutive pages and is required to predict the action that navigates from the previous page to the current one. This approach enables the model to better capture the relationships between UI states and the corresponding user actions.

\subsubsection{Incorporating Reasoning into the Action Task}
\label{Incorporating Reasoning into the Action Task}
While action annotations provide direct supervision, they often omit the intermediate reasoning required to bridge natural language instructions and executable actions. To address this gap, we incorporate a reasoning process into the action task, enabling models to better infer intent and align actions with GUI context.

\textbf{Rejection Sampling for Enhanced Reasoning.} Based on large scale open-source and self-collected action data within a unified action space, we employ rejection sampling ~\citep{deepseek_r1} to systematically generate reasoning trajectories. By Leveraging VLMs (e.g., UI-TARS~\citep{UI-TARS} and Qwen2.5VL~\citep{qwen2.5vl}) with GUI reasoning capabilities, We repeatedly sample the thought process of action selection, and quantitatively compare the sampled actions with the ground-truth action labels to compute the sampling success rate. By imposing a threshold on the success rate, we are able to filter and curate large-scale, high-quality datasets of action reasoning instances.

\textbf{Planning-oriented Reasoning.} 
In high-level task scenarios, reasoning plays a crucial role in enabling models to make coherent and accurate decisions across extended action sequences. Prior work demonstrates the benefits of integrating reasoning into large language model (LLM) decision-making. For example, the ReAct framework interleaves reasoning and acting to enable interactive agents \citep{ReAct}. Additionally, recent advances emphasize the importance of key reasoning components such as self-reflection and planning for efficient task execution \citep{reflexion, dagan2023dynamicplanningllm, manus}.

Motivated by these insights, we design our reasoning data to incorporate these principles within the context of GUI task execution, adopting a \textbf{planning-oriented reasoning format} in which each reasoning step simultaneously involves both planning and action execution. Specifically, at every step, the model observes environmental information, including the current GUI page, action history, and the current plan, to produce an updated plan along with the next action to execute.

This reasoning process includes two complementary components:
\begin{itemize}
  \item \textbf{Plan formulation and adjustment}: updating or confirming the plan based on the current context.
  \item \textbf{Action execution}: determining and performing the next action according to the plan.
\end{itemize}

This design balances the need for accurate planning with execution efficiency by integrating plan updates and action decisions within the same reasoning step.

\subsection{Data Refinement}
Although the annotated dataset provides a strong foundation, further refinement is necessary to ensure training effectiveness. In particular, we focus on two key aspects: data selection, which filters and balances samples to enhance coverage across tasks and screen scenarios, and data mixing strategies, which control the composition of training batches based on task categories and image resolutions. These refinements enhance the overall data distribution and assist the model in generalizing to varied GUI environments.

\subsubsection{Data Selection}
We employ a greedy and diversity-oriented data selection strategy called NovelSelect~\citep{yang2025measuring} to iteratively select the most "novel" samples and create a diverse training dataset. Taking both inter-sample distance and uneven information density into consideration, NovelSelect defines the "novelty" of a new sample $x$ relative to an existing dataset $\mathcal{Z}$:
\begin{align}
v(x) = \sum_{\substack{x_j \in \mathcal{Z}}} w(x, x_j)^\alpha \cdot \sigma(x_j)^\beta \cdot d(x, x_j),
\end{align}
where $w(x, x_j)$ is defined as the function of the rank of $x_j$ in the sorted list of distances from $x$ to all points in $\mathcal{Z}$, $\sigma(x_j)$ is a density factor reflecting the average distance to the $K$-nearest neighbors of $x_j$, and $d(\cdot,~\cdot)$ donates the distance between the embeddings of samples. $\alpha$ and $\beta$ are hyperparameters, set to 1.0 and 0.5, respectively.

In our work, we entend NovelSelect from text-only approach to multimodal dataset. We calculat the embeddings of samples by concatenating the embeddings of images and corresponding text. The combined embedding from multimodal data is utilized to calculate the "novelty" score for each sample in the target dataset. At each iteration, the sample with the highest score is selected into $\mathcal{Z}$. This process continued until the data budget is exhausted. We achieve the same performance as training with the full dataset using only 50\% of the data by this method.

\subsubsection{Data Mixing}
Data mixing is conducted along two dimensions: category-based mixing by tasks and resolution-based mixing by images. This approach helps maintain a balanced distribution of functional and visual complexity in training batches.
\paragraph{Category-Based Data Mixing}
The mixture of open-source and proprietary data significantly impacts training efficiency and model performance~\citep{ye2024data, dong2023abilities, liu2024regmix, zhang2024mm1}, particularly in scenarios where balancing the injection of domain-specific knowledge and the retention of general knowledge is crucial. Based on the finding that the influence of different data categories on themselves is much greater than on other categories~\citep{zhang2024mm1}, we adopt a straightforward yet effective experimental approach to determine the appropriate mixing ratios for different data categories. First, we categorize GUI-related tasks into five categories: widget understanding (grounding and referring), widget-related VQA, single-step operation, information extraction, and captioning. For each data category, we gradually increase the amount of training data until the model achieved our expected performance in that specific category. Finally, we mix the different data categories according to the data volumes determined in the previous steps to train our final model.

\paragraph{Resolution-Based Data Mixing}
When training multimodal models, it is common to impose a maximum resolution limit~\citep{Qwen2-VL}. Given that the majority of our proprietary data comprises high-resolution mobile screenshots, while real-world applications may encounter images of varying resolutions, it is imperative for the model to learn to perform tasks effectively across different resolutions. To address this, we employ a dynamic resolution training strategy. Specifically, for each training batch, a maximum resolution value, \textit{max\_pixels}, is randomly sampled from a normal distribution. Subsequently, each image within the batch is randomly assigned a final resolution, \textit{training\_pixels}, which does not exceed \textit{max\_pixels}. The distribution of \textit{training\_pixels} is designed to follow the pixel distribution observed in general datasets. Finally, the model is trained under various \textit{max\_pixels} settings to determine the final dynamic resolution scheme.

\subsection{Data Statistics}
Following the processes of low-quality data filtering and proportional adjustment of diversity in data categories and resolutions, the total amount of data used in Continued Pre-training (CPT) is 7.8 million, comprising 3.1 million in-house data, 2.8 million open-source general data, and 1.9 million open-source GUI navigation data. Moreover, Reinforcement Fine-tuning (RFT) trains only on the subsets of GUI navigation action data. Among our in-house data, element referring data, element grounding data, element description, screen caption data, screen VQA data, and navigation action data account for 17.2\%, 24.1\%, 10.2\%, 23\%, 11.8\%, and 13.7\%, respectively.
\section{Experiments}

\begin{table}[!t]
\centering
\caption{Training parameters for the MagicGUI model.}
\label{tab:cpt}
\renewcommand{\arraystretch}{1.35}
\begin{tabular}{lcl}
\toprule[1.5pt]
\textbf{Parameter} & \textbf{Default Value} & \textbf{Description} \\
\midrule[1pt]
max\_prompt\_length & 8192 & Maximum prompt length \\
max\_new\_token & 2048 & Maximum completion length \\
max\_pixels & 602112 & Maximum pixels  \\
num\_train\_epochs & 1 & Number of training epochs \\
weight\_decay & 0.1 & Weight decay coefficient \\
adam\_beta1 & 0.9 & Adam optimizer beta1 parameter \\
adam\_beta2 & 0.95 & Adam optimizer beta2 parameter \\
max\_grad\_norm & 0.5 & Maximum gradient norm for clipping \\
lr\_scheduler\_type & cosine & Learning rate scheduler type \\
bf16 & True & Use bfloat16 precision \\
\midrule[1pt]
\multicolumn{3}{l}{\textbf{Continue Pre-training (CPT)}} \\
num\_gpus & 64 & Number of GPUs (8 machines × 8 GPUs) \\
per\_device\_train\_batch\_size & 2 & Training batch size per device \\
gradient\_accumulation\_steps & 4 & Gradient accumulation steps \\
foundational\_learning\_rate & 2e-5 & Learning rate of Foundational Pre-training \\
annealing\_learning\_rate & 1e-5 & Learning rate of Annealing \\
\midrule[1pt]
\multicolumn{3}{l}{\textbf{Reinforcement Fine-tuning (RFT)}} \\
num\_gpus & 32 & Number of GPUs (4 machines × 8 GPUs) \\
per\_device\_train\_batch\_size & 8 & Training batch size per device \\
gradient\_accumulation\_steps & 1 & Gradient accumulation steps \\
learning\_rate & 1e-6 & Learning rate of actor policy \\
num\_generations & 8 & Number of generations in DF-GRPO \\
temperature & 0.9 & Temperature parameter for generations in DF-GRPO \\
top\_p & 0.96 & Cumulative probability threshold for generation \\
top\_k & 20 & Number of top tokens filtered by highest probability \\
\bottomrule[1.5pt]
\end{tabular}
\end{table}

We compare MagicGUI with various baselines, including commercial models such as GPT-4o, Gemini-2.0-flash, as well as academic models like Qwen series models Qwen2-VL-7B~\citep{Qwen2-VL} Qwen2.5-VL (7B and 32B)~\citep{qwen2.5vl}, UI-TARS~\citep{UI-TARS} series models UI-TARS-7B and UI-TARS-1.5-7B, the MiMo model MiMo-VL-7B~\citep{coreteam2025mimovltechnicalreport} and AgentCPM-GUI~\citep{AgentCPM-GUI}. The training details of our MagicGUI are listed in Table~\ref{tab:cpt}. In addition, we investigate the effect of explicit reasoning in MagicGUI in Appendix~\ref{ablation}.

\subsection{Test Datasets}
\label{testdataset}
We employ a series of test datasets, including both open-source and proprietary datasets, to evaluate the capabilities of MagicGUI. Datasets are categorized into two primary groups: perception, and GUI agent, which are listed in Table~\ref{tab:my_label}.

\textbf{Perception.} Perception capability is typically assessed through Visual Question Answering (VQA) and grounding. The open-source VQA benchmark \textit{ScreenQA-short} \citep{hsiao2024screenqa} evaluates web structural comprehension and mobile screen content understanding through QA tasks. For grounding capability, We focus on two open-source benchmarks: \textit{ScreenSpot v2} \citep{Os-atlas} and \textit{Os-Atlas-mobile} \citep{Os-atlas}. These benchmarks assess the ability to understand and localize elements within GUIs, with all annotations providing UI bounding boxes. They evaluate GUI grounding across mobile, desktop, and web platforms. Given our primary focus on the model's performance exclusively on mobile devices (including phones and tablets), we extract the relevant test data for mobile platforms. 

\textbf{GUI Agent.} 
To evaluate the ultimate GUI Agent Capability of MagicGUI, we conduct experiments on two open-source benchmarks: \textit{AndroidControl} \citep{ac} and \textit{GUI-Odyssey} \citep{odyssey}. AndroidControl assesses planning and action-execution capabilities in mobile environments. This dataset comprises two task categories: (1) high-level tasks requiring autonomous multi-step planning and execution by GUI agent models, and (2) low-level tasks involving step-wise instruction. GUI-Odyssey, a specialized benchmark for cross-application navigation in mobile interfaces, is distinguished by complex workflows averaging over 15 steps per task.

Moreover, we have identified three significant shortcomings in these two open-source test datasets. First, domestic apps differ substantially from those in the international market, yet both Android Control and GUI Odyssey are based on international market apps. Second, these datasets represent the ground-truth positions of UI controls solely by coordinate points rather than bounding boxes, complicating the evaluation of whether the model's inferred results are truly accurate. For example, in GUI-Odyssey, a distance threshold of less than 14\% of the screenshot width between coordinate points is used to assess the correctness of the model's inferred grounding. Finally, these datasets are constructed under ideal conditions, resulting in insufficient coverage of exceptional scenarios.

Therefore, we construct the \textbf{\textit{Magic-RICH}} test dataset based on real usage scenarios, encompassing \underline{\textbf{R}}outine, \underline{\textbf{I}}nstruction, \underline{\textbf{C}}omplex, and \underline{\textbf{H}}andling exception scenarios. This dataset covers 17 categories and over 150 apps that are highly popular in the domestic market, with each subset comprising 1,000 samples. Furthermore, we incorporate special operations, such as screenshots and long screenshots, which are not covered in previous datasets. Routine includes high-frequency operations extract from actual navigation tasks, while the difference between Instruction and complex mainly lies in the way of questioning. Direct and simple questioning methods are categorized in the Instruction dataset, while relatively more complex questions (logical thinking, graphical judgment, intentional instructions, etc.) are placed in the complex dataset. For the exception scenarios, we define conditions such as non-interactive, completed, and loading, and conducted targeted data collection to effectively evaluate the model's capabilities in practical applications. 

\begin{table}[!t]
    \centering
    \caption{Information for all test datasets.}
    \renewcommand{\arraystretch}{1.3}
    \begin{tabular}{llcc}
    \toprule[1.5pt]
       Data Set Name & Test Capability & Items & Language \\
       \midrule[1pt]
       \multicolumn{3}{l}{\textit{\textbf{Perception Capability}}} \\
       ScreenQA-short \citep{hsiao2024screenqa} & GUI VQA & 8427 & EN \\
       ScreenSpot v2 mobile \citep{Os-atlas} & GUI grounding & 501 & EN\\
       Os-Atlas-mobile \citep{Os-atlas} & GUI grounding & 503 & EN \\
       \midrule[1pt]
       \multicolumn{3}{l}{\textit{\textbf{GUI Agent Capability}}}  \\
       Android Control \citep{ac}  & High-level and Low-level tasks & 7836 & EN\\
       GUI-Odyssey \citep{odyssey}  &  High-level task & 22927 & EN \\
       \rowcolor{blue!10} Magic-RICH (self-build)  &  Low-level task & 4000 & CN \\
       
    \bottomrule[1.5pt]

    \end{tabular}
    \label{tab:my_label}
\end{table}

\begin{table}[!t]
\caption{For referring benchmark. We report the average scores from various GUI agent models. For grounding benchmarks. We report the accuracy of output point within the bounding box. \textbf{Bold} and \underline{underline} indicate the best and second-best results.}
\label{tab:referring-grounding}
\centering
  \renewcommand{\arraystretch}{1.3}
  \begin{tabular}{lccc}
  \toprule[1.5pt]
  {\textbf{Agent Models}}  & {\textbf{ScreenQA-short}} & {\textbf{ScreenSpot v2 mobile}} & {\textbf{Os-Atlas-mobile}}  \\
  
  \midrule[1pt]
  \multicolumn{4}{l}{\textit{\textbf{Closed-source Models}}}  \\
  GPT-4o~\citep{hurst2024gpt}  & 90.3 &  10.6  &  4.6  \\
  Gemini 2.0~\citep{Gemini2.0}  & 90.4 & 10.6 & 5.8 \\
  \midrule[1pt]
  \multicolumn{4}{l}{\textit{\textbf{Open-source Models}}}  \\
  InternVL-2-8B~\citep{chen2024internvl}  & 88.4  & 4.2  & 2.4   \\
  Qwen2-VL-7B~\citep{Qwen2-VL}   & 92.6  & 70.7  &  27.2   \\
  Qwen2.5-VL-7B~\citep{qwen2.5vl}    & 92.1 & 56.1  &  26.6  \\
  UI-TARS-7B~\citep{UI-TARS}   & \textbf{95.4} & \underline{88.6} & \underline{82.5}  \\
  UI-TARS-1.5-7B~\citep{ui-tars-15-seed}  & 93.0 & 85.8 & 79.3   \\
  \midrule[1pt]
  \rowcolor{blue!10} MagicGUI-CPT  & \underline{94.6} & \textbf{90.2} & \textbf{95.2}  \\
  \bottomrule[1.5pt]
  \end{tabular}
\end{table}

\subsection{Perception Capability}

For the evaluation of VQA tasks, we establish five levels based on the corresponding judgment criteria regarding the completeness and reasonableness of the answers. We utilize the Qwen2.5-7B-Instruct\citep{qwen2.5} model for evaluation in a few-shot manner, wherein a template comprises three distinct inference results and the ground truth for each level. For questions with multiple answers, we select the highest score as the final score for that question. Based on the experimental results, MagicGUI-CPT demonstrates performance comparable to the state-of-the-art UI-TARS. Specifically, it achieves an average score of 94.6, as illustrated in Table~\ref{tab:referring-grounding}. This result reflects MagicGUI-CPT’s enhanced proficiency in understanding and interpreting UI components with high precision and comprehensiveness.

Regarding grounding capability, various metrics are available for evaluation, including the distance between two center points, Intersection over Union (IoU), and point inclusion within the bounding box. Given the necessity for the agent to interact with selectable UI elements in practical GUI applications, we ultimately select point inclusion within the bounding box as the criterion for assessing grounding capabilities. Furthermore, for models that output only bounding boxes, we employ the center point of the bounding box for evaluation.

As demonstrated in Table~\ref{tab:referring-grounding}, MagicGUI-CPT exhibits superior capability in accurately localizing UI elements within bounding boxes on mobile platforms, achieving 90.2\% and 95.2\% accuracy in ScreenSpot v2 mobile and Os-Atlas-mobile, respectively. This performance underscores the model’s robustness and generalization across diverse datasets. The adoption of point inclusion within bounding boxes as the evaluation metric aligns with practical application requirements, further validating MagicGUI-CPT’s effectiveness in real-world GUI grounding tasks.

\begin{table}[!t]
\caption{Performance comparison on the Magic-RICH dataset. We report the accuracy of action type (Type), grounding (Grd), and step success rate (SR). \textbf{Bold} and \underline{underline} indicate the best and second-best results.}
\label{tab:Honor-dataset}
\centering
  \renewcommand{\arraystretch}{1.3}
  \begin{tabular}{lcccccccccc}
  \toprule[1.5pt]
  \multirow{2}{*}{\textbf{Agent Models}} & \multicolumn{3}{c}{\textbf{Routine}} & \multicolumn{3}{c}{\textbf{Instruction}} & \multicolumn{3}{c}{\textbf{Complex}} & \textbf{Handling} \\
  \cmidrule(lr){2-4} \cmidrule(lr){5-7} \cmidrule(lr){8-10}  & \small{Type} & \small{Grd} & \small{SR}  & \small{Type} & \small{Grd} & \small{SR} & \small{Type} & \small{Grd} & \small{SR} & \textbf{Exception} \\
  \midrule[1pt]
  \multicolumn{10}{l}{\textit{\textbf{Closed-source Models}}}  \\
  GPT-4o~\citep{hurst2024gpt} & 49.3 &  16.7  & 4.6  & 56.6 & 13.5 & 19.8 &  49.0 & 14.6  & 7.4 & 85.1   \\
  Gemini 2.0~\citep{Gemini2.0} &  89.2 & 49.4 & 34.7  & 84.1 & 54.2 & 51.4  & 83.3  & 50.3 & 42.0 & 73.7  \\
  \midrule[1pt]
  \multicolumn{10}{l}{\textit{\textbf{Open-source Models}}}  \\
  InternVL-2-8B~\citep{chen2024internvl}  & 30.1  & 2.8  &  1.3 & 37.1  & 4.0 & 15.8 & 17.1 & 6.0  & 1.3 & 70.8  \\
  Qwen2-VL-7B~\citep{Qwen2-VL}   & 71.7  & 41.0  & 28.1 & 73.6  & 43.9  & 41.5 &  65.6 & 28.7  & 21.2 & 68.3 \\
  Qwen2.5-VL-7B~\citep{qwen2.5vl}  & 94.3  & 92.6  & 76.3 & 89.3  & \underline{95.7} & 83.6 &  86.6 & 69.6 & 60.0 & 67.0   \\
  UI-TARS-7B~\citep{UI-TARS} & 83.5 & 84.9  & 73.3 & 76.6  & 85.6 & 69.8 & 91.4 & 69.1  & 67.0 & 3.6   \\
  UI-TARS-1.5-7B~\citep{ui-tars-15-seed}  & 85.6  & 96.2  & 81.5 & 78.6  & 92.1  & 72.2 &  \textbf{94.7} & 74.3  & 71.1 & 1.0   \\
  MiMo-VL-7B-SFT~\citep{coreteam2025mimovltechnicalreport}  & 93.0  & 77.9  & 65.3 & 89.7  & 85.7  & 75.4 &  89.1 & 80.1  & 71.0 & 57.0 \\
  AgentCPM-GUI~\citep{AgentCPM-GUI}  & 84.3  & 92.2  & 75.1 & 70.4 & 80.7 & 56.0 &  72.3 & 54.6  & 39.4 & 2.4   \\
  \midrule[1pt]
  \rowcolor{blue!10} MagicGUI-CPT & \underline{98.5}  & \textbf{98.5}  & \underline{97.2} & \underline{95.5}  & \textbf{96.3}  & \underline{92.9} & 88.5 & \textbf{82.3}  & \underline{72.9} & \textbf{93.2}   \\
  \rowcolor{blue!10} MagicGUI-RFT & \textbf{99.7}  & \underline{97.5}  & \textbf{97.5} & \textbf{97.2}  & 95.6 & \textbf{94.0} &  \underline{92.1} & \underline{80.4} & \textbf{74.1} & \underline{92.1} \\
  \bottomrule[1.5pt]
  \end{tabular}
\end{table}

\begin{table}[!t]
\caption{Performance comparison on open-source AndroidControl and GUI-Odyssey datasets. We report the action type accuracy (Type) and step success rate (SR). \textbf{Bold} and \underline{underline} indicate the best and second-best results. *OS-Atlas uses different train/test splits on GUI-Odyssey benchmark and is not directly comparable.}
\label{tab:Open-dataset}
   \centering
  \renewcommand{\arraystretch}{1.3}
  \begin{tabular}{l > {\centering\arraybackslash}p{1.1cm} >{\centering\arraybackslash}p{1.1cm} >{\centering\arraybackslash}p{1.1cm} >{\centering\arraybackslash}p{1.1cm} >{\centering\arraybackslash}p{1.1cm}  >{\centering\arraybackslash}p{1.1cm}} 
  \toprule[1.5pt]
  \multirow{2}{*}{\textbf{Agent Models}} & \multicolumn{2}{c}{\textbf{AC-Low}} & \multicolumn{2}{c}{\textbf{AC-High}} & \multicolumn{2}{c}{\textbf{GUI-Odyssey}} \\
  \cmidrule(lr){2-3} \cmidrule(lr){4-5} \cmidrule(lr){6-7}
  & Type & SR & Type & SR  & Type & SR \\
  \midrule[1pt]
  \multicolumn{5}{l}{\textit{\textbf{Closed-source Models}}}  \\
  GPT-4o~\citep{hurst2024gpt}  & -  &  19.5 &  -  & 20.8  & -  & 20.4  \\
  
  Gemini 2.0~\citep{Gemini2.0} & - & 28.5 &  - & 60.2   & -  & 3.3 \\
  
  Claude 2.0~\citep{Claude2.0} & - & 28.5 &  - & 12.5   & 60.9  & -   \\
  \midrule[1pt]
  \multicolumn{5}{l}{\textit{\textbf{Open-source Models}}}  \\

  Qwen2-VL-7B~\citep{Qwen2-VL}  & 55.7 & 36.2 & 45.8 & 21.2 & 58.6 & 13.3\\
  
  Qwen2.5-VL-7B~\citep{qwen2.5vl}  & 94.1 & 85.0 & 75.1 & 62.9 & 59.5 & 46.3 \\
  
  Aguvis-7B~\citep{Aguvis}  &  93.9 &  89.4 &  65.6 & 54.2 & 26.7 & 13.5  \\
 
  OS-Atlas-7B~\citep{Os-atlas} & 73.0 & 67.3 & 70.4 & 56.5 & 91.8* & 76.8*  \\

  UI-TARS-7B~\citep{UI-TARS}  & \underline{95.2} & \underline{91.8} & 81.6 & \underline{74.4} & 86.1 & 67.9  \\

  AgentCPM-GUI~\citep{AgentCPM-GUI} & 94.4 & 90.2 & 77.7  & 69.2 & \textbf{90.9} & \textbf{75.0}  \\
  \midrule[1pt]
  \rowcolor{blue!10} MagicGUI-CPT & 94.5 & 86.7 & \underline{84.6} & 73.1 & \underline{90.4} & 73.5 \\
  \rowcolor{blue!10} MagicGUI-RFT & \textbf{97.2} &  \textbf{93.5} &  \textbf{84.7} & \textbf{76.3}  & 89.7 & \underline{74.3} \\
  \bottomrule[1.5pt]
  \end{tabular}
\end{table}

\subsection{GUI Agent Capability}

To comprehensively assess the capabilities of the MagicGUI Agent, our evaluation primarily focuses on the Magic-RICH datasets alongside open-source benchmarks. The Magic-RICH datasets are specifically curated to address the Chinese language context and business-related scenarios, ensuring relevance and applicability to the target HONOR mobile devices. In this benchmark, we design a unified evaluation prompt template, except for UI-TARS and AgentCPM-GUI, which utilize their own prompts, as demonstrated in Appendix~\ref{prompt}.

The performance comparison between the MagicGUI agent and other baselines on the Magic-RICH dataset is presented in Table~\ref{tab:Honor-dataset}. For each subset, we report action type accuracy (Type), grounding accuracy (Grd), and step success rate (SR), where Type solely assesses the accuracy of action types, Grd evaluates the grounding accuracy, and SR additionally considers all parameters. MagicGUI-RFT achieves the highest SR of approximately 97.5\%, 94.0\%, and 74.1\% on the Routine, Instruction, and Complex subsets, respectively. We observe that RFT may decrease the performance of CPT on the Handling exception subset; we suspect that this is the cost of enhancing generalization performance, which may confuse certain normal and exceptional examples. Furthermore, our spatially enhanced composite reward function only considers the accuracy of coordinates for samples with correct action types to prevent the phenomenon of reward hacking. Therefore, the overall grounding capability of RFT shows a decline in metrics, while achieving a higher SR.

To evaluate the generalization performance of our MagicGUI, we conduct a series of comprehensive experiments on open-source benchmarks: AndroidControl and GUI-Odyssey. Specifically, the AndroidControl benchmark comprises AndroidControl-Low and AndroidControl-High versions, where AndroidControl-Low provides step-level instructions, whereas AndroidControl-High contains only task-level queries, thereby posing a greater challenge to GUI agent models. We follow the AgentCPM-GUI~\citep{AgentCPM-GUI} to ensure a fair comparison among different baselines, noting that OS-Atlas employs a different train/test split on the GUI-Odyssey benchmark. Given that our tap and text input actions include coordinate position information, and their logic does not align with that of the open-source benchmarks, we have adjusted the criteria for these two actions during evaluation. Additionally, for the Navigate back action, our model may directly output a tap action to click on the back arrow on the screen, and the relevant detection criteria have also been modified. As shown in Table \ref{tab:Open-dataset}, MagicGUI-RFT achieves the highest step success rates (SR) of approximately 93.5\% and 76.3\% on the AndroidControl-Low and AndroidControl-High benchmarks, respectively. On the GUI-Odyssey benchmark, MagicGUI-RFT achieves performance comparable to the state-of-the-art AgentCPM-GUI, reflecting the remarkable generalization performance of MagicGUI.
\section{Future Work}
The current model primarily relies on CPT's robust low-level operation accuracy and the generalization improvements brought by RFT, leading to strong performance on the Magic-RICH benchmark. However, due to the limited availability of high-level Chinese app tasks in our training dataset, we will gradually establish a pipeline for constructing these training and testing data, and this work continuously improves. In addition, we will focus on developing more universal and efficient commercial applications for HONOR mobile devices. The following aspects will be considered.

\textbf{Comprehensive Multimodal Model.} 
Future research will aim to develop a more comprehensive and unified multimodal model that integrates text, images, speech and videos inputs to enhance the model’s understanding and reasoning capabilities across complex multimodal data. Furthermore, investigating more efficient interaction and fusion mechanisms among different modalities will improve the model’s generalization and applicability in real-world scenarios.

\textbf{User Interaction.}
Enhancing the model’s interaction capabilities with users, especially in understanding and responding to natural language or speech instructions within GUI, represents  a promising direction. Future research may explore more flexible and natural interaction methods, including multi-turn dialogues, proactive clarification queries, and personalized interaction strategies to improve user experience and task execution accuracy.

\textbf{Memory and Personalization.}
Future research could investigate incorporating long-term memory mechanisms that enable the model to retain user preferences, historical operations, and contextual information, thereby facilitating personalized task execution and recommendations. Additionally, dynamic memory updating and management techniques could support continual learning and adaptation across diverse user tasks and scenarios.

\textbf{Edge-Cloud Collaboration.}
Considering the computational limitations of mobile devices, future work may investigate architectures for edge-cloud collaboration that dynamically allocate computing tasks and optimize resource utilization. This approach can enhance real-time responsiveness and privacy protection while ensuring efficient execution of complex long-horizontal tasks.

\textbf{Tool Call and MCP Service.}
Exploring the model’s capability to invoke external tools and Model Context Protocol (MCP) services is another critical direction. This includes integrating HONOR Native agents and third-party APIs, automating script execution, and enabling cross-platform data exchange to expand the model’s functional capabilities and enhance its practical flexibility in real-world applications.

\section{Conclusion}
This paper presents MagicGUI, a foundational mobile GUI agent designed to address perception, grounding, and reasoning in real-world GUI environments. We introduce the GUI Data Pipeline to construct the largest and most diverse GUI-centric multimodal dataset to date, and propose a unified annotation framework covering both perception and interaction tasks. Our two-stage training paradigm enables robust and generalizable learning across a wide range of benchmarks. These advancements collectively support the system-level usage scenarios demonstrated in Appendix~\ref{system-level}. In future work, we aim to extend MagicGUI with multi-turn user interaction capability, richer multimodal inputs, user personalization, and support for real-world deployment through tool invocation and edge-cloud collaboration.

\bibliographystyle{delta_tuning}
\bibliography{references}

\begin{thebibliography}{71}
\providecommand{\natexlab}[1]{#1}
\providecommand{\url}[1]{\texttt{#1}}
\expandafter\ifx\csname urlstyle\endcsname\relax
  \providecommand{\doi}[1]{doi: #1}\else
  \providecommand{\doi}{doi: \begingroup \urlstyle{rm}\Url}\fi

\bibitem[Anthropic(2024)]{Claude2.0}
Anthropic.
\newblock Introducing computer use, a new claude 3.5 sonnet, and claude 3.5 haiku., 2024.
\newblock URL \url{https://www.anthropic.com/news/3-5-models-and-computer-use, 2024}.

\bibitem[Bai et~al.(2024)Bai, Zhou, Pan, Cemri, Suhr, Levine, and Kumar]{digirl}
Hao Bai, Yifei Zhou, Jiayi Pan, Mert Cemri, Alane Suhr, Sergey Levine, and Aviral Kumar.
\newblock Digirl: Training in-the-wild device-control agents with autonomous reinforcement learning.
\newblock \emph{Advances in Neural Information Processing Systems (NeurIPS)}, 37:\penalty0 12461--12495, 2024.

\bibitem[Bai et~al.(2025)Bai, Chen, Liu, Wang, Ge, Song, Dang, Wang, Wang, Tang, et~al.]{qwen2.5vl}
Shuai Bai, Keqin Chen, Xuejing Liu, Jialin Wang, Wenbin Ge, Sibo Song, Kai Dang, Peng Wang, Shijie Wang, Jun Tang, et~al.
\newblock Qwen2.5-vl technical report.
\newblock \emph{arXiv preprint arXiv:2502.13923}, 2025.

\bibitem[Chai et~al.(2024)Chai, Huang, Niu, Xiao, Liu, Zhang, Gao, Ren, and Li]{amex}
Yuxiang Chai, Siyuan Huang, Yazhe Niu, Han Xiao, Liang Liu, Dingyu Zhang, Peng Gao, Shuai Ren, and Hongsheng Li.
\newblock Amex: Android multi-annotation expo dataset for mobile gui agents.
\newblock \emph{arXiv preprint arXiv:2407.17490}, 2024.

\bibitem[Chen et~al.(2024{\natexlab{a}})Chen, Cui, Hu, Qin, Fang, Zhao, Wang, Liu, Chen, Huo, et~al.]{guiact}
Wentong Chen, Junbo Cui, Jinyi Hu, Yujia Qin, Junjie Fang, Yue Zhao, Chongyi Wang, Jun Liu, Guirong Chen, Yupeng Huo, et~al.
\newblock Guicourse: From general vision language models to versatile gui agents.
\newblock \emph{arXiv preprint arXiv:2406.11317}, 2024{\natexlab{a}}.

\bibitem[Chen et~al.(2024{\natexlab{b}})Chen, Wang, Tian, Ye, Gao, Cui, Tong, Hu, Luo, Ma, et~al.]{chen2024far}
Zhe Chen, Weiyun Wang, Hao Tian, Shenglong Ye, Zhangwei Gao, Erfei Cui, Wenwen Tong, Kongzhi Hu, Jiapeng Luo, Zheng Ma, et~al.
\newblock How far are we to gpt-4v? closing the gap to commercial multimodal models with open-source suites.
\newblock \emph{Science China Information Sciences}, 67\penalty0 (12):\penalty0 220101, 2024{\natexlab{b}}.

\bibitem[Chen et~al.(2024{\natexlab{c}})Chen, Wu, Wang, Su, Chen, Xing, Zhong, Zhang, Zhu, Lu, et~al.]{chen2024internvl}
Zhe Chen, Jiannan Wu, Wenhai Wang, Weijie Su, Guo Chen, Sen Xing, Muyan Zhong, Qinglong Zhang, Xizhou Zhu, Lewei Lu, et~al.
\newblock Internvl: Scaling up vision foundation models and aligning for generic visual-linguistic tasks.
\newblock In \emph{IEEE/CVF Conference on Computer Vision and Pattern Recognition (CVPR)}, pp.\  24185--24198, 2024{\natexlab{c}}.

\bibitem[Cheng et~al.(2024{\natexlab{a}})Cheng, Guan, Wu, and Yan]{cheng2024least}
Chuanqi Cheng, Jian Guan, Wei Wu, and Rui Yan.
\newblock From the least to the most: Building a plug-and-play visual reasoner via data synthesis.
\newblock \emph{arXiv preprint arXiv:2406.19934}, 2024{\natexlab{a}}.

\bibitem[Cheng et~al.(2024{\natexlab{b}})Cheng, Sun, Chu, Xu, Li, Zhang, and Wu]{seeclick_screenspot}
Kanzhi Cheng, Qiushi Sun, Yougang Chu, Fangzhi Xu, Yantao Li, Jianbing Zhang, and Zhiyong Wu.
\newblock Seeclick: Harnessing gui grounding for advanced visual gui agents.
\newblock \emph{arXiv preprint arXiv:2401.10935}, 2024{\natexlab{b}}.

\bibitem[Dagan et~al.(2023)Dagan, Keller, and Lascarides]{dagan2023dynamicplanningllm}
Gautier Dagan, Frank Keller, and Alex Lascarides.
\newblock Dynamic planning with a llm, 2023.
\newblock URL \url{https://arxiv.org/abs/2308.06391}.

\bibitem[Dong et~al.(2023)Dong, Yuan, Lu, Li, Xue, Liu, Wang, Yuan, Zhou, and Zhou]{dong2023abilities}
Guanting Dong, Hongyi Yuan, Keming Lu, Chengpeng Li, Mingfeng Xue, Dayiheng Liu, Wei Wang, Zheng Yuan, Chang Zhou, and Jingren Zhou.
\newblock How abilities in large language models are affected by supervised fine-tuning data composition.
\newblock \emph{arXiv preprint arXiv:2310.05492}, 2023.

\bibitem[Dou et~al.(2024)Dou, Liu, Jia, Zhou, Xiong, Shan, Huang, Wang, Fan, Xi, et~al.]{stepcoder}
Shihan Dou, Yan Liu, Haoxiang Jia, Enyu Zhou, Limao Xiong, Junjie Shan, Caishuang Huang, Xiao Wang, Xiaoran Fan, Zhiheng Xi, et~al.
\newblock Stepcoder: Improving code generation with reinforcement learning from compiler feedback.
\newblock In \emph{Proceedings of the Annual Meeting of the Association for Computational Linguistics (ACL)}, pp.\  4571--4585, 2024.

\bibitem[Gou et~al.(2025)Gou, Wang, Zheng, Xie, Chang, Shu, Sun, and Su]{UGround}
Boyu Gou, Ruohan Wang, Boyuan Zheng, Yanan Xie, Cheng Chang, Yiheng Shu, Huan Sun, and Yu~Su.
\newblock Navigating the digital world as humans do: Universal visual grounding for gui agents.
\newblock In \emph{International Conference on Learning Representations (ICLR)}, 2025.

\bibitem[Grattafiori et~al.(2024)Grattafiori, Dubey, Jauhri, Pandey, Kadian, Al-Dahle, Letman, Mathur, Schelten, Vaughan, et~al.]{grattafiori2024llama}
Aaron Grattafiori, Abhimanyu Dubey, Abhinav Jauhri, Abhinav Pandey, Abhishek Kadian, Ahmad Al-Dahle, Aiesha Letman, Akhil Mathur, Alan Schelten, Alex Vaughan, et~al.
\newblock The llama 3 herd of models.
\newblock \emph{arXiv preprint arXiv:2407.21783}, 2024.

\bibitem[Gu et~al.(2024)Gu, Zhang, Zhou, Yu, Xing, Wang, Cao, Jia, Zhang, Wang, et~al.]{gu2024infinity}
Shuhao Gu, Jialing Zhang, Siyuan Zhou, Kevin Yu, Zhaohu Xing, Liangdong Wang, Zhou Cao, Jintao Jia, Zhuoyi Zhang, Yixuan Wang, et~al.
\newblock Infinity-mm: Scaling multimodal performance with large-scale and high-quality instruction data.
\newblock \emph{arXiv preprint arXiv:2410.18558}, 2024.

\bibitem[Guo et~al.(2025)Guo, Yang, Zhang, Song, Zhang, Xu, Zhu, Ma, Wang, Bi, et~al.]{deepseek_r1}
Daya Guo, Dejian Yang, Haowei Zhang, Junxiao Song, Ruoyu Zhang, Runxin Xu, Qihao Zhu, Shirong Ma, Peiyi Wang, Xiao Bi, et~al.
\newblock Deepseek-r1: Incentivizing reasoning capability in llms via reinforcement learning.
\newblock \emph{arXiv preprint arXiv:2501.12948}, 2025.

\bibitem[Hong et~al.(2024)Hong, Wang, Lv, Xu, Yu, Ji, Wang, Wang, Dong, Ding, et~al.]{cogagent}
Wenyi Hong, Weihan Wang, Qingsong Lv, Jiazheng Xu, Wenmeng Yu, Junhui Ji, Yan Wang, Zihan Wang, Yuxiao Dong, Ming Ding, et~al.
\newblock Cogagent: A visual language model for gui agents.
\newblock In \emph{IEEE/CVF Conference on Computer Vision and Pattern Recognition (CVPR)}, pp.\  14281--14290, 2024.

\bibitem[Hsiao et~al.(2024)Hsiao, Zubach, Wang, and Chen]{hsiao2024screenqa}
Yu-Chung Hsiao, Fedir Zubach, Maria Wang, and Jindong Chen.
\newblock Screenqa: Large-scale question-answer pairs over mobile app screenshots, 2024.

\bibitem[Hurst et~al.(2024)Hurst, Lerer, Goucher, Perelman, Ramesh, Clark, Ostrow, Welihinda, Hayes, Radford, et~al.]{hurst2024gpt}
Aaron Hurst, Adam Lerer, Adam~P Goucher, Adam Perelman, Aditya Ramesh, Aidan Clark, AJ~Ostrow, Akila Welihinda, Alan Hayes, Alec Radford, et~al.
\newblock Gpt-4o system card.
\newblock \emph{arXiv preprint arXiv:2410.21276}, 2024.

\bibitem[Jaech et~al.(2024)Jaech, Kalai, Lerer, Richardson, El-Kishky, Low, Helyar, Madry, Beutel, Carney, et~al.]{openai_o1}
Aaron Jaech, Adam Kalai, Adam Lerer, Adam Richardson, Ahmed El-Kishky, Aiden Low, Alec Helyar, Aleksander Madry, Alex Beutel, Alex Carney, et~al.
\newblock Openai o1 system card.
\newblock \emph{arXiv preprint arXiv:2412.16720}, 2024.

\bibitem[Li et~al.(2024{\natexlab{a}})Li, Fang, Smyrnis, Ivgi, Jordan, Gadre, Bansal, Guha, Keh, Arora, et~al.]{li2024datacomp}
Jeffrey Li, Alex Fang, Georgios Smyrnis, Maor Ivgi, Matt Jordan, Samir~Yitzhak Gadre, Hritik Bansal, Etash Guha, Sedrick~Scott Keh, Kushal Arora, et~al.
\newblock Datacomp-lm: In search of the next generation of training sets for language models.
\newblock \emph{Advances in Neural Information Processing Systems (NeurIPS)}, 37:\penalty0 14200--14282, 2024{\natexlab{a}}.

\bibitem[Li et~al.(2025)Li, Meng, Lin, Luo, Tian, Ma, Huang, and Chua]{screenspot_pro}
Kaixin Li, Ziyang Meng, Hongzhan Lin, Ziyang Luo, Yuchen Tian, Jing Ma, Zhiyong Huang, and Tat-Seng Chua.
\newblock Screenspot-pro: Gui grounding for professional high-resolution computer use.
\newblock \emph{arXiv preprint arXiv:2504.07981}, 2025.

\bibitem[Li et~al.(2024{\natexlab{b}})Li, Bishop, Li, Rawles, Campbell-Ajala, Tyamagundlu, and Riva]{ac}
Wei Li, William~E Bishop, Alice Li, Christopher Rawles, Folawiyo Campbell-Ajala, Divya Tyamagundlu, and Oriana Riva.
\newblock On the effects of data scale on ui control agents.
\newblock \emph{Advances in Neural Information Processing Systems (NeurIPS)}, 37:\penalty0 92130--92154, 2024{\natexlab{b}}.

\bibitem[Li et~al.(2020)Li, Li, He, Zheng, Li, and Guan]{li2020widget}
Yang Li, Gang Li, Luheng He, Jingjie Zheng, Hong Li, and Zhiwei Guan.
\newblock Widget captioning: Generating natural language description for mobile user interface elements.
\newblock \emph{arXiv preprint arXiv:2010.04295}, 2020.

\bibitem[Lin et~al.(2025)Lin, Li, Gao, Yang, Wu, Bai, Lei, Wang, and Shou]{Showui}
Kevin~Qinghong Lin, Linjie Li, Difei Gao, Zhengyuan Yang, Shiwei Wu, Zechen Bai, Stan~Weixian Lei, Lijuan Wang, and Mike~Zheng Shou.
\newblock Showui: One vision-language-action model for gui visual agent.
\newblock In \emph{IEEE/CVF Conference on Computer Vision and Pattern Recognition (CVPR)}, pp.\  19498--19508, 2025.

\bibitem[Liu et~al.(2024)Liu, Zheng, Muennighoff, Zeng, Dou, Pang, Jiang, and Lin]{liu2024regmix}
Qian Liu, Xiaosen Zheng, Niklas Muennighoff, Guangtao Zeng, Longxu Dou, Tianyu Pang, Jing Jiang, and Min Lin.
\newblock Regmix: Data mixture as regression for language model pre-training.
\newblock \emph{arXiv preprint arXiv:2407.01492}, 2024.

\bibitem[Liu et~al.(2025{\natexlab{a}})Liu, Li, Xie, Hu, Han, Zhang, Yang, and Wu]{infigui-r1}
Yuhang Liu, Pengxiang Li, Congkai Xie, Xavier Hu, Xiaotian Han, Shengyu Zhang, Hongxia Yang, and Fei Wu.
\newblock Infigui-r1: Advancing multimodal gui agents from reactive actors to deliberative reasoners.
\newblock \emph{arXiv preprint arXiv:2504.14239}, 2025{\natexlab{a}}.

\bibitem[Liu et~al.(2025{\natexlab{b}})Liu, Zhang, Liu, Zhang, Sun, and Wang]{othink_mr1}
Zhiyuan Liu, Yuting Zhang, Feng Liu, Changwang Zhang, Ying Sun, and Jun Wang.
\newblock Othink-mr1: Stimulating multimodal generalized reasoning capabilities via dynamic reinforcement learning.
\newblock \emph{arXiv preprint arXiv:2503.16081}, 2025{\natexlab{b}}.

\bibitem[Liu et~al.(2025{\natexlab{c}})Liu, Sun, Zang, Dong, Cao, Duan, Lin, and Wang]{visual_rft}
Ziyu Liu, Zeyi Sun, Yuhang Zang, Xiaoyi Dong, Yuhang Cao, Haodong Duan, Dahua Lin, and Jiaqi Wang.
\newblock Visual-rft: Visual reinforcement fine-tuning.
\newblock \emph{arXiv preprint arXiv:2503.01785}, 2025{\natexlab{c}}.

\bibitem[Lu et~al.(2024{\natexlab{a}})Lu, Shao, Liu, Meng, Li, Chen, Huang, Zhang, Qiao, and Luo]{odyssey}
Quanfeng Lu, Wenqi Shao, Zitao Liu, Fanqing Meng, Boxuan Li, Botong Chen, Siyuan Huang, Kaipeng Zhang, Yu~Qiao, and Ping Luo.
\newblock Gui odyssey: A comprehensive dataset for cross-app gui navigation on mobile devices.
\newblock \emph{arXiv preprint arXiv:2406.08451}, 2024{\natexlab{a}}.

\bibitem[Lu et~al.(2024{\natexlab{b}})Lu, Yang, Shen, and Awadallah]{lu2024omniparser}
Yadong Lu, Jianwei Yang, Yelong Shen, and Ahmed Awadallah.
\newblock Omniparser for pure vision based gui agent.
\newblock \emph{arXiv preprint arXiv:2408.00203}, 2024{\natexlab{b}}.

\bibitem[Lu et~al.(2025)Lu, Chai, Guo, Yin, Liu, Wang, Xiao, Ren, Xiong, and Li]{ui_r1}
Zhengxi Lu, Yuxiang Chai, Yaxuan Guo, Xi~Yin, Liang Liu, Hao Wang, Han Xiao, Shuai Ren, Guanjing Xiong, and Hongsheng Li.
\newblock Ui-r1: Enhancing action prediction of gui agents by reinforcement learning.
\newblock \emph{arXiv preprint arXiv:2503.21620}, 2025.

\bibitem[Luo et~al.(2025)Luo, Wang, He, and Xia]{gui_r1}
Run Luo, Lu~Wang, Wanwei He, and Xiaobo Xia.
\newblock Gui-r1: A generalist r1-style vision-language action model for gui agents.
\newblock \emph{arXiv preprint arXiv:2504.10458}, 2025.

\bibitem[Nguyen et~al.(2024)Nguyen, Chen, Wang, Wu, Park, Hu, Lyu, Wu, Aponte, Xia, et~al.]{nguyen2024gui}
Dang Nguyen, Jian Chen, Yu~Wang, Gang Wu, Namyong Park, Zhengmian Hu, Hanjia Lyu, Junda Wu, Ryan Aponte, Yu~Xia, et~al.
\newblock Gui agents: A survey.
\newblock \emph{arXiv preprint arXiv:2412.13501}, 2024.

\bibitem[Pichai et~al.(2024)Pichai, Hassabis, and Kavukcuoglu]{Gemini2.0}
Sundar Pichai, D~Hassabis, and K~Kavukcuoglu.
\newblock Introducing gemini 2.0: our new ai model for the agentic era, 2024.

\bibitem[Qin et~al.(2025)Qin, Ye, Fang, Wang, Liang, Tian, Zhang, Li, Li, Huang, et~al.]{UI-TARS}
Yujia Qin, Yining Ye, Junjie Fang, Haoming Wang, Shihao Liang, Shizuo Tian, Junda Zhang, Jiahao Li, Yunxin Li, Shijue Huang, et~al.
\newblock Ui-tars: Pioneering automated gui interaction with native agents.
\newblock \emph{arXiv preprint arXiv:2501.12326}, 2025.

\bibitem[Schneider et~al.(2024)Schneider, Meske, and Kuss]{schneider2024foundation}
Johannes Schneider, Christian Meske, and Pauline Kuss.
\newblock Foundation models: a new paradigm for artificial intelligence.
\newblock \emph{Business \& Information Systems Engineering}, 66\penalty0 (2):\penalty0 221--231, 2024.

\bibitem[Seed(2025)]{ui-tars-15-seed}
ByteDance Seed.
\newblock Ui-tars-1.5.
\newblock \url{https://seed-tars.com/1.5}, 2025.

\bibitem[Shao et~al.(2024)Shao, Wang, Zhu, Xu, Song, Bi, Zhang, Zhang, Li, Wu, et~al.]{deepseek_math}
Zhihong Shao, Peiyi Wang, Qihao Zhu, Runxin Xu, Junxiao Song, Xiao Bi, Haowei Zhang, Mingchuan Zhang, YK~Li, Y~Wu, et~al.
\newblock Deepseekmath: Pushing the limits of mathematical reasoning in open language models.
\newblock \emph{arXiv preprint arXiv:2402.03300}, 2024.

\bibitem[Shen et~al.(2025)Shen, Liu, Li, Fang, Ma, Liao, Shen, Zhang, Zhao, Zhang, et~al.]{vlmr1}
Haozhan Shen, Peng Liu, Jingcheng Li, Chunxin Fang, Yibo Ma, Jiajia Liao, Qiaoli Shen, Zilun Zhang, Kangjia Zhao, Qianqian Zhang, et~al.
\newblock Vlm-r1: A stable and generalizable r1-style large vision-language model.
\newblock \emph{arXiv preprint arXiv:2504.07615}, 2025.

\bibitem[Shen \& Yang(2025)Shen and Yang]{manus}
Minjie Shen and Qikai Yang.
\newblock From mind to machine: The rise of manus ai as a fully autonomous digital agent, 2025.
\newblock URL \url{https://arxiv.org/abs/2505.02024}.

\bibitem[Shinn et~al.(2023)Shinn, Cassano, Berman, Gopinath, Narasimhan, and Yao]{reflexion}
Noah Shinn, Federico Cassano, Edward Berman, Ashwin Gopinath, Karthik Narasimhan, and Shunyu Yao.
\newblock Reflexion: Language agents with verbal reinforcement learning, 2023.
\newblock URL \url{https://arxiv.org/abs/2303.11366}.

\bibitem[Wang et~al.(2024{\natexlab{a}})Wang, Xu, Ye, Yan, Shen, Zhang, Huang, and Sang]{mobileagent}
Junyang Wang, Haiyang Xu, Jiabo Ye, Ming Yan, Weizhou Shen, Ji~Zhang, Fei Huang, and Jitao Sang.
\newblock Mobile-agent: Autonomous multi-modal mobile device agent with visual perception.
\newblock \emph{arXiv preprint arXiv:2401.16158}, 2024{\natexlab{a}}.

\bibitem[Wang et~al.(2024{\natexlab{b}})Wang, Ma, Feng, Zhang, Yang, Zhang, Chen, Tang, Chen, Lin, et~al.]{wang2024survey}
Lei Wang, Chen Ma, Xueyang Feng, Zeyu Zhang, Hao Yang, Jingsen Zhang, Zhiyuan Chen, Jiakai Tang, Xu~Chen, Yankai Lin, et~al.
\newblock A survey on large language model based autonomous agents.
\newblock \emph{Frontiers of Computer Science}, 18\penalty0 (6):\penalty0 186345, 2024{\natexlab{b}}.

\bibitem[Wang et~al.(2024{\natexlab{c}})Wang, Bai, Tan, Wang, Fan, Bai, Chen, Liu, Wang, Ge, Fan, Dang, Du, Ren, Men, Liu, Zhou, Zhou, and Lin]{Qwen2-VL}
Peng Wang, Shuai Bai, Sinan Tan, Shijie Wang, Zhihao Fan, Jinze Bai, Keqin Chen, Xuejing Liu, Jialin Wang, Wenbin Ge, Yang Fan, Kai Dang, Mengfei Du, Xuancheng Ren, Rui Men, Dayiheng Liu, Chang Zhou, Jingren Zhou, and Junyang Lin.
\newblock Qwen2-vl: Enhancing vision-language model's perception of the world at any resolution.
\newblock \emph{arXiv preprint arXiv:2409.12191}, 2024{\natexlab{c}}.

\bibitem[Wang et~al.(2024{\natexlab{d}})Wang, Wu, Liu, Hao, Wang, and Shao]{distrl}
Taiyi Wang, Zhihao Wu, Jianheng Liu, Jianye Hao, Jun Wang, and Kun Shao.
\newblock Distrl: An asynchronous distributed reinforcement learning framework for on-device control agents.
\newblock \emph{arXiv preprint arXiv:2410.14803}, 2024{\natexlab{d}}.

\bibitem[Wang et~al.(2024{\natexlab{e}})Wang, Lv, Yu, Hong, Qi, Wang, Ji, Yang, Zhao, XiXuan, et~al.]{Cogvlm}
Weihan Wang, Qingsong Lv, Wenmeng Yu, Wenyi Hong, Ji~Qi, Yan Wang, Junhui Ji, Zhuoyi Yang, Lei Zhao, Song XiXuan, et~al.
\newblock Cogvlm: Visual expert for pretrained language models.
\newblock \emph{Advances in Neural Information Processing Systems (NeurIPS)}, 37:\penalty0 121475--121499, 2024{\natexlab{e}}.

\bibitem[Wang \& Liu(2024)Wang and Liu]{wang2024oscar}
Xiaoqiang Wang and Bang Liu.
\newblock Oscar: Operating system control via state-aware reasoning and re-planning.
\newblock \emph{arXiv preprint arXiv:2410.18963}, 2024.

\bibitem[Wu et~al.(2025{\natexlab{a}})Wu, Ma, Wang, Yu, Lu, and Liu]{wu2025gui}
Penghao Wu, Shengnan Ma, Bo~Wang, Jiaheng Yu, Lewei Lu, and Ziwei Liu.
\newblock Gui-reflection: Empowering multimodal gui models with self-reflection behavior.
\newblock \emph{arXiv preprint arXiv:2506.08012}, 2025{\natexlab{a}}.

\bibitem[Wu et~al.(2024{\natexlab{a}})Wu, Xu, Liu, Tan, Liu, Li, Luan, Wang, and Shang]{mobilevlm}
Qinzhuo Wu, Weikai Xu, Wei Liu, Tao Tan, Jianfeng Liu, Ang Li, Jian Luan, Bin Wang, and Shuo Shang.
\newblock Mobilevlm: A vision-language model for better intra- and inter-ui understanding, 2024{\natexlab{a}}.
\newblock URL \url{https://arxiv.org/abs/2409.14818}.

\bibitem[Wu et~al.(2025{\natexlab{b}})Wu, Liu, Luan, and Wang]{reachagent}
Qinzhuo Wu, Wei Liu, Jian Luan, and Bin Wang.
\newblock Reachagent: Enhancing mobile agent via page reaching and operation.
\newblock \emph{arXiv preprint arXiv:2502.02955}, 2025{\natexlab{b}}.

\bibitem[Wu et~al.(2024{\natexlab{b}})Wu, Han, Ding, Weng, Liu, Yao, Yu, and Kong]{wu2024copilot}
Zhiyong Wu, Chengcheng Han, Zichen Ding, Zhenmin Weng, Zhoumianze Liu, Shunyu Yao, Tao Yu, and Lingpeng Kong.
\newblock Os-copilot: Towards generalist computer agents with self-improvement.
\newblock \emph{arXiv preprint arXiv:2402.07456}, 2024{\natexlab{b}}.

\bibitem[Wu et~al.(2024{\natexlab{c}})Wu, Wu, Xu, Wang, Sun, Jia, Cheng, Ding, Chen, Liang, et~al.]{Os-atlas}
Zhiyong Wu, Zhenyu Wu, Fangzhi Xu, Yian Wang, Qiushi Sun, Chengyou Jia, Kanzhi Cheng, Zichen Ding, Liheng Chen, Paul~Pu Liang, et~al.
\newblock Os-atlas: A foundation action model for generalist gui agents.
\newblock \emph{arXiv preprint arXiv:2410.23218}, 2024{\natexlab{c}}.

\bibitem[Xia et~al.(2024)Xia, Deng, Dunn, and Zhang]{xia2024agentless}
Chunqiu~Steven Xia, Yinlin Deng, Soren Dunn, and Lingming Zhang.
\newblock Agentless: Demystifying llm-based software engineering agents.
\newblock \emph{arXiv preprint arXiv:2407.01489}, 2024.

\bibitem[Xiaomi(2025)]{coreteam2025mimovltechnicalreport}
LLM-Core-Team Xiaomi.
\newblock Mimo-vl technical report, 2025.
\newblock URL \url{https://arxiv.org/abs/2506.03569}.

\bibitem[Xie et~al.(2025)Xie, Gao, Ren, Luo, Hong, Dai, Zhou, Qiu, Wu, and Luo]{logic_rl}
Tian Xie, Zitian Gao, Qingnan Ren, Haoming Luo, Yuqian Hong, Bryan Dai, Joey Zhou, Kai Qiu, Zhirong Wu, and Chong Luo.
\newblock Logic-rl: Unleashing llm reasoning with rule-based reinforcement learning.
\newblock \emph{arXiv preprint arXiv:2502.14768}, 2025.

\bibitem[Xie et~al.(2024)Xie, Zhang, Chen, Li, Zhao, Cao, Hua, Cheng, Shin, Lei, et~al.]{xie2024osworld}
Tianbao Xie, Danyang Zhang, Jixuan Chen, Xiaochuan Li, Siheng Zhao, Ruisheng Cao, Toh~J Hua, Zhoujun Cheng, Dongchan Shin, Fangyu Lei, et~al.
\newblock Osworld: Benchmarking multimodal agents for open-ended tasks in real computer environments.
\newblock \emph{Advances in Neural Information Processing Systems}, 37:\penalty0 52040--52094, 2024.

\bibitem[Xu et~al.(2024)Xu, Wang, Wang, Lu, Xie, Saha, Sahoo, Yu, and Xiong]{Aguvis}
Yiheng Xu, Zekun Wang, Junli Wang, Dunjie Lu, Tianbao Xie, Amrita Saha, Doyen Sahoo, Tao Yu, and Caiming Xiong.
\newblock Aguvis: Unified pure vision agents for autonomous gui interaction.
\newblock \emph{arXiv preprint arXiv:2412.04454}, 2024.

\bibitem[Yang et~al.(2024{\natexlab{a}})Yang, Yang, Zhang, Hui, Zheng, Yu, Li, Liu, Huang, Wei, Lin, Yang, Tu, Zhang, Yang, Yang, Zhou, Lin, Dang, Lu, Bao, Yang, Yu, Li, Xue, Zhang, Zhu, Men, Lin, Li, Xia, Ren, Ren, Fan, Su, Zhang, Wan, Liu, Cui, Zhang, and Qiu]{qwen2.5}
An~Yang, Baosong Yang, Beichen Zhang, Binyuan Hui, Bo~Zheng, Bowen Yu, Chengyuan Li, Dayiheng Liu, Fei Huang, Haoran Wei, Huan Lin, Jian Yang, Jianhong Tu, Jianwei Zhang, Jianxin Yang, Jiaxi Yang, Jingren Zhou, Junyang Lin, Kai Dang, Keming Lu, Keqin Bao, Kexin Yang, Le~Yu, Mei Li, Mingfeng Xue, Pei Zhang, Qin Zhu, Rui Men, Runji Lin, Tianhao Li, Tingyu Xia, Xingzhang Ren, Xuancheng Ren, Yang Fan, Yang Su, Yichang Zhang, Yu~Wan, Yuqiong Liu, Zeyu Cui, Zhenru Zhang, and Zihan Qiu.
\newblock Qwen2.5 technical report.
\newblock \emph{arXiv preprint arXiv:2412.15115}, 2024{\natexlab{a}}.

\bibitem[Yang et~al.(2025{\natexlab{a}})Yang, Li, Yang, Zhang, Hui, Zheng, Yu, Gao, Huang, Lv, Zheng, Liu, Zhou, Huang, Hu, Ge, Wei, Lin, Tang, Yang, Tu, Zhang, Yang, Yang, Zhou, Zhou, Lin, Dang, Bao, Yang, Yu, Deng, Li, Xue, Li, Zhang, Wang, Zhu, Men, Gao, Liu, Luo, Li, Tang, Yin, Ren, Wang, Zhang, Ren, Fan, Su, Zhang, Zhang, Wan, Liu, Wang, Cui, Zhang, Zhou, and Qiu]{qwen3tech}
An~Yang, Anfeng Li, Baosong Yang, Beichen Zhang, Binyuan Hui, Bo~Zheng, Bowen Yu, Chang Gao, Chengen Huang, Chenxu Lv, Chujie Zheng, Dayiheng Liu, Fan Zhou, Fei Huang, Feng Hu, Hao Ge, Haoran Wei, Huan Lin, Jialong Tang, Jian Yang, Jianhong Tu, Jianwei Zhang, Jianxin Yang, Jiaxi Yang, Jing Zhou, Jingren Zhou, Junyang Lin, Kai Dang, Keqin Bao, Kexin Yang, Le~Yu, Lianghao Deng, Mei Li, Mingfeng Xue, Mingze Li, Pei Zhang, Peng Wang, Qin Zhu, Rui Men, Ruize Gao, Shixuan Liu, Shuang Luo, Tianhao Li, Tianyi Tang, Wenbiao Yin, Xingzhang Ren, Xinyu Wang, Xinyu Zhang, Xuancheng Ren, Yang Fan, Yang Su, Yichang Zhang, Yinger Zhang, Yu~Wan, Yuqiong Liu, Zekun Wang, Zeyu Cui, Zhenru Zhang, Zhipeng Zhou, and Zihan Qiu.
\newblock Qwen3 technical report, 2025{\natexlab{a}}.
\newblock URL \url{https://arxiv.org/abs/2505.09388}.

\bibitem[Yang et~al.(2023)Yang, Zhang, Li, Zou, Li, and Gao]{yang2023set}
Jianwei Yang, Hao Zhang, Feng Li, Xueyan Zou, Chunyuan Li, and Jianfeng Gao.
\newblock Set-of-mark prompting unleashes extraordinary visual grounding in gpt-4v.
\newblock \emph{arXiv preprint arXiv:2310.11441}, 2023.

\bibitem[Yang et~al.(2024{\natexlab{b}})Yang, Wang, Li, Luo, Chen, Huang, and Li]{Aria-UI}
Yuhao Yang, Yue Wang, Dongxu Li, Ziyang Luo, Bei Chen, Chao Huang, and Junnan Li.
\newblock Aria-ui: Visual grounding for gui instructions.
\newblock \emph{arXiv preprint arXiv:2412.16256}, 2024{\natexlab{b}}.

\bibitem[Yang et~al.(2025{\natexlab{b}})Yang, Nan, Ye, Dou, Wang, Li, Lv, Wu, Gui, Zhang, et~al.]{yang2025measuring}
Yuming Yang, Yang Nan, Junjie Ye, Shihan Dou, Xiao Wang, Shuo Li, Huijie Lv, Mingqi Wu, Tao Gui, Qi~Zhang, et~al.
\newblock Measuring data diversity for instruction tuning: A systematic analysis and a reliable metric.
\newblock \emph{arXiv preprint arXiv:2502.17184}, 2025{\natexlab{b}}.

\bibitem[Yao et~al.(2023)Yao, Zhao, Yu, Du, Shafran, Narasimhan, and Cao]{ReAct}
Shunyu Yao, Jeffrey Zhao, Dian Yu, Nan Du, Izhak Shafran, Karthik Narasimhan, and Yuan Cao.
\newblock React: Synergizing reasoning and acting in language models, 2023.
\newblock URL \url{https://arxiv.org/abs/2210.03629}.

\bibitem[Ye et~al.(2024)Ye, Liu, Sun, Zhan, Zhou, and Qiu]{ye2024data}
Jiasheng Ye, Peiju Liu, Tianxiang Sun, Jun Zhan, Yunhua Zhou, and Xipeng Qiu.
\newblock Data mixing laws: Optimizing data mixtures by predicting language modeling performance.
\newblock \emph{arXiv preprint arXiv:2403.16952}, 2024.

\bibitem[You et~al.(2024)You, Zhang, Schoop, Weers, Swearngin, Nichols, Yang, and Gan]{you2024ferret}
Keen You, Haotian Zhang, Eldon Schoop, Floris Weers, Amanda Swearngin, Jeffrey Nichols, Yinfei Yang, and Zhe Gan.
\newblock Ferret-ui: Grounded mobile ui understanding with multimodal llms.
\newblock In \emph{European Conference on Computer Vision (ECCV)}, pp.\  240--255. Springer, 2024.

\bibitem[Zhang et~al.(2024{\natexlab{a}})Zhang, Li, He, Zhang, Qiao, Qin, Ma, Kang, Lin, Rajmohan, et~al.]{zhang2024ufo}
Chaoyun Zhang, Liqun Li, Shilin He, Xu~Zhang, Bo~Qiao, Si~Qin, Minghua Ma, Yu~Kang, Qingwei Lin, Saravan Rajmohan, et~al.
\newblock Ufo: A ui-focused agent for windows os interaction.
\newblock \emph{arXiv preprint arXiv:2402.07939}, 2024{\natexlab{a}}.

\bibitem[Zhang et~al.(2025{\natexlab{a}})Zhang, Yang, Liu, Li, Han, Chen, Huang, Fu, and Yu]{appagent}
Chi Zhang, Zhao Yang, Jiaxuan Liu, Yanda Li, Yucheng Han, Xin Chen, Zebiao Huang, Bin Fu, and Gang Yu.
\newblock Appagent: Multimodal agents as smartphone users.
\newblock In \emph{Proceedings of the CHI conference on Human Factors in Computing Systems (CHI)}, pp.\  1--20, 2025{\natexlab{a}}.

\bibitem[Zhang et~al.(2024{\natexlab{b}})Zhang, Gao, Gan, Dufter, Wenzel, Huang, Shah, Du, Zhang, Li, et~al.]{zhang2024mm1}
Haotian Zhang, Mingfei Gao, Zhe Gan, Philipp Dufter, Nina Wenzel, Forrest Huang, Dhruti Shah, Xianzhi Du, Bowen Zhang, Yanghao Li, et~al.
\newblock Mm1. 5: Methods, analysis \& insights from multimodal llm fine-tuning.
\newblock \emph{arXiv preprint arXiv:2409.20566}, 2024{\natexlab{b}}.

\bibitem[Zhang et~al.(2025{\natexlab{b}})Zhang, Lu, Fu, Huo, Yang, Wu, Si, Cong, Chen, Lin, et~al.]{AgentCPM-GUI}
Zhong Zhang, Yaxi Lu, Yikun Fu, Yupeng Huo, Shenzhi Yang, Yesai Wu, Han Si, Xin Cong, Haotian Chen, Yankai Lin, et~al.
\newblock Agentcpm-gui: Building mobile-use agents with reinforcement fine-tuning.
\newblock \emph{arXiv preprint arXiv:2506.01391}, 2025{\natexlab{b}}.

\bibitem[Zhou et~al.(2025)Zhou, Dai, Wang, Zhou, Jia, et~al.]{gui_g1}
Yuqi Zhou, Sunhao Dai, Shuai Wang, Kaiwen Zhou, Qinqlin Jia, et~al.
\newblock Gui-g1: Understanding r1-zero-like training for visual grounding in gui agents.
\newblock \emph{arXiv preprint arXiv:2505.15810}, 2025.

\end{thebibliography}

\newpage
\appendix
\section{Reasoning Ablation}
\label{ablation}
\textbf{Reasoning on Fundamental Click Actions.} To investigate the effect of explicit reasoning in basic GUI operations, we focus on single-step click actions, which represent the most fundamental and frequent type of user interaction. This experiment is designed to evaluate the model's ability to process complex user interface information and make accurate decisions under varying instruction styles. We sample a subset from our Magic-RICH test dataset, as described in Section~\ref{testdataset}, containing only tap actions. This subset covers three categories of queries: direct instructions, indirect instructions, and spatial instructions, where direct instructions originate from direct description scenarios, and the latter two come from complex description scenarios.

We first prepared a foundation model (\texttt{CPT-no-act}), trained solely on auxiliary data (such as VQA and grounding data) without explicit action supervision, following the CPT training setup in Section~\ref{First Stage}. Next, we constructed an action dataset consisting of 30k samples, including medium-level and state-transition actions. Following the reasoning data synthesis approach described previously, we generated a dataset in which half of the samples were annotated with explicit step-by-step reasoning traces. We used this reasoning-augmented subset to train the model \texttt{CPT-half-act-think}. Subsequently, we further improved the model by applying DF-GRPO training for the remaining half data, resulting in \texttt{CPT-full-act-think}. For comparison, we also trained a model (\texttt{CPT-full-act}) using the same total 30k action data without any reasoning traces. One should note that the reasoning trace for single-step action typically includes the analysis of the given instruction and a detailed description of current GUI interface.

\begin{figure}[!ht]
    \centering
    \includegraphics[scale=0.5]{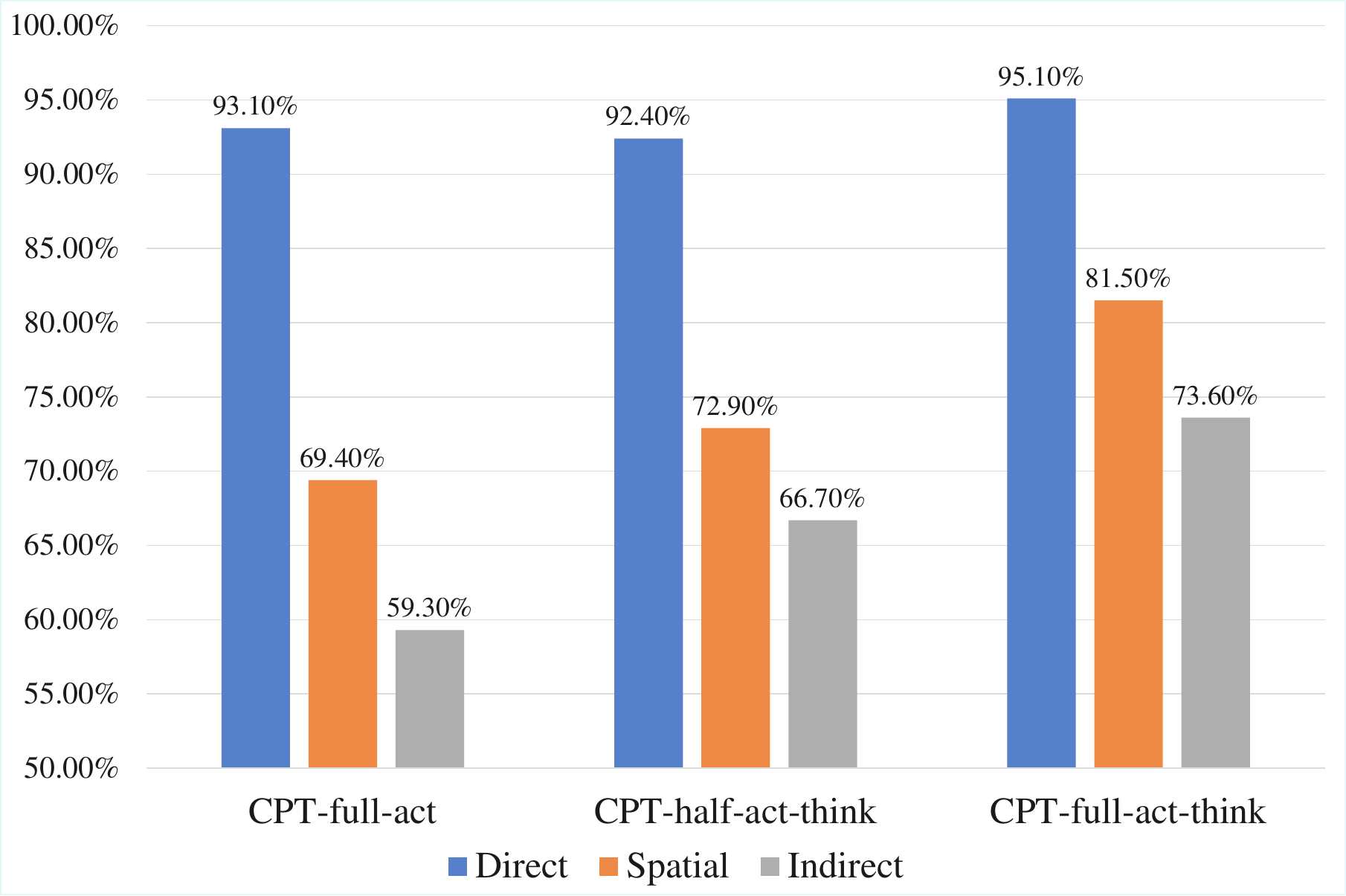}
    \caption{Ablation results on direct, spatial, and indirect instructions for different training setups.}
    \label{fig:click-reasoning-results}
\end{figure}

\textbf{Experimental Results.} As shown in Figure~\ref{fig:click-reasoning-results}, explicit reasoning brings significant improvements. Notably, \texttt{CPT-half-act-think}, trained with only half of the data annotated with reasoning traces, already surpasses the fully supervised \texttt{CPT-full-act} on indirect and spatial instructions, while achieving comparable performance on direct instructions. Furthermore, after applying RL training on the reasoning-augmented model, the performance is further enhanced, especially for indirect and spatial instructions. Specifically, the accuracy for direct instructions improves from \textbf{93.1\%} to \textbf{95.1\%}, for spatial instructions from \textbf{69.4\%} to \textbf{81.5\%}, and for indirect instructions from \textbf{59.3\%} to \textbf{73.6\%}. These results demonstrate that injecting explicit reasoning ability, combined with RL optimization, substantially boosts the model's generalization and robustness on fundamental GUI action tasks. In particular, the \texttt{think}-augmented model exhibits a clear advantage on challenging cases involving spatial and indirect instructions.

\section{Prompt Templates}
\label{prompt}
\subsection{MagicGUI}
\begin{tcolorbox}[breakable, colback=black!5!white,colframe=black!75!black,title=MagicGUI: A Unified Prompt Template]
\tcbsubtitle{System Message}
\begin{CJK}{UTF8}{gbsn}
你是一个手机移动助手。
\end{CJK}

\tcbsubtitle{User}
\begin{CJK}{UTF8}{gbsn}
已知用户在界面<image>，提出了要求:

[user\_request] \\

你认为合理的单步操作是什么？除了函数调用之外，你不能输出任何其他内容。你可以调用以下函数来控制智能手机：\\

UI基础操作：

\quad 1.tap(x,y)该函数用于在智能手机屏幕上点击特定点，坐标 x 和 y 表示待点击控件中心位置。
  
\quad 2.scroll(x,y,direction) 该函数用于从起始坐标 (x,y) 开始在智能手机屏幕上滑动操作，direction为手指滑动的方向，可以是"up"、"down"、"left" 或 "right"。
  
\quad 3.text(x,y,text\_input) 该函数用于在智能手机屏幕上输入指定的文本text\_input。坐标 x 和 y 表示待点击控件的中心位置。 \\

手机按键操作：

\quad 4.navigate\_back() 该函数用于返回智能手机的上一个屏幕。

\quad 5.navigate\_home() 该函数用于返回手机的home screen。 \\

其他操作：

\quad 6.long\_press(x,y) 该函数用于在智能手机屏幕上的特定点执行长按操作。坐标 x 和 y 表示待点击控件的中心位置。

\quad 7.wait() 该函数表示在当前页面等候。

\quad 8.enter() 该函数表示按下enter键。

\quad 9.take\_over(message) 该函数用于提示用户接管智能手机，其中 message 是提示用户接管手机的原因。如果原因不确定，请填写“请您接管当前界面”。

\quad 10.drag(x1,y1,x2,y2) 该函数执行一个对起始和终点敏感的拖动操作，表示手指从点(x1,y1)拖到点(x2,y2)。常见的场景包括滑块拖动、滚动选择器拖动和图片裁剪。

\quad 11.screen\_shot() 该函数用于截图。

\quad 12.long\_screen\_shot() 该函数用于长截图。

\quad 13.call\_api(api\_name,operation) 对指定的APP进行操作。api\_name是API的名称。operation可以选择open或者kill。例如，call\_api(Amazon, open)意味着打开亚马逊APP。

如果你发现当前指令无法在当前页面上执行，你需要输出no\_answer()。如果你发现当前指令已完成，你需要输出action\_completed()。
\end{CJK}

[current\_screenshot]

\tcbsubtitle{Assistant}
[thought\_and\_action]
\end{tcolorbox}

\subsection{UI-TARS}

\begin{tcolorbox}[breakable, colback=black!5!white,colframe=black!75!black,title=UI-TARS Prompt Template]
\tcbsubtitle{System Message}
You are a helpful assistant.
\tcbsubtitle{User}
You are a GUI agent. You are given a task and your action history, with screenshots. You need to perform the next action to complete the task.\\

\textbf{Output Format}\\

{Thought: \ldots}\\
{Action: \ldots}\\

\vspace{1em}
\textbf{Action Space}\\
{click(start\_box=\textquotesingle\textless|box\_start|\textgreater(x1,y1)\textless|box\_end|\textgreater\textquotesingle)}\\
{long\_press(start\_box=\textquotesingle\textless|box\_start|\textgreater(x1,y1)\textless|box\_end|\textgreater\textquotesingle, time=\textquotesingle \textquotesingle)}\\
{type(content=\textquotesingle \textquotesingle)  \#If you want to submit your input, use "\textbackslash n" at the end of content. }\\
{scroll(direction=\textquotesingle down or up or right or left\textquotesingle)}\\
{open\_app(app\_name=\textquotesingle \textquotesingle)} \\
{drag(start\_box=\textquotesingle<|box\_start|>(x1,y1)<|box\_end|>\textquotesingle, \\ end\_box=\textquotesingle<|box\_start|>(x2,y2)<|box\_end|>\textquotesingle)} \\
{press\_back()}\\
{press\_home()}\\
{no\_answer()}\\
{wait()}\\
{action\_completed() \# Use escape characters \textquotesingle, ", and \textbackslash n in content part to ensure we can parse the content in normal python string format.}\\
{finished()} \# Submit the task regardless of whether it succeeds or fails.\\

\vspace{1em}
\textbf{Note}\\
- Use English in Thought part.\\
- Summarize your next action (with its target element) in one sentence in Thought part.\\

\textbf{User Instruction}\\
{[user\_request]}

\tcbsubtitle{Assistant}
Thought: [low\_lew\_instruction]

Action:

\end{tcolorbox}

\subsection{AgentCPM-GUI}
\begin{tcolorbox}[breakable, colback=black!5!white,colframe=black!75!black,title=AgentCPM-GUI Prompt Template]
\tcbsubtitle{System Message}
\begin{CJK}{UTF8}{gbsn}
你是一个手机移动助手。
\end{CJK}

\tcbsubtitle{User}
\begin{CJK}{UTF8}{gbsn}
\# Role

\quad 你是一名熟悉安卓系统触屏GUI操作的智能体，将根据用户的问题，分析当前界面的GUI元素和布局，生成相应的操作。 \\

\# Task

\quad 针对用户问题，根据输入的当前屏幕截图，输出下一步的操作。 \\

\# Rule
\quad  - 以紧凑JSON格式输出

\quad  - 输出操作必须遵循Schema约束 \\

\# Schema

\{

"type":"object",

"description":"执行操作并决定当前任务状态",

"additionalProperties":false,

"required": ["thought"],

"properties":

    \quad \{"thought":
    
    \quad \quad \{"type":"string","description":"智能体的思维过程"\},
    
    \quad \quad "POINT":\{"ref":"/defs/Location","description":"点击屏幕上的指定位置"\},
    
    \quad \quad "to":
        \{"description":"移动，组合手势参数",
        
    \quad \quad "oneOf":
          [\{"enum["up","down","left","right"],
          
    \quad \quad "description":"从当前点出发，执行滑动手势操作，方向包括向上、向下、向左、向右"\},
    
    \quad \quad \{"ref":"/defs/Location","description":"移动到某个位置"\}]\},
    
    \quad \quad "duration":{"type":"integer","description":"动作执行的时间或等待时间，毫秒",
    
    \quad \quad "minimum":0,"default":200},
    
    \quad \quad "PRESS":\{"type":"string",
    
    \quad \quad "description":"触发特殊按键，HOME为回到主页按钮，BACK为返回按钮，ENTER为回车按钮",
          
    \quad \quad "enum":["HOME","BACK","ENTER"]\},
    
    \quad \quad "TYPE":{"type":"string","description":"输入文本"},
    
    \quad \quad "STATUS":\{"type":"string",
    
    \quad \quad "description":"当前任务的状态。
    
    \quad \quad \quad \quad 特殊情况：satisfied，
    
    \quad \quad \quad \quad 无需操作；impossible，
    
    \quad \quad \quad \quad 任务无法完成；
    
    \quad \quad \quad \quad interrupt，任务中断；
    
    \quad \quad \quad \quad need\_feedback，需要用户反馈；",
          
    \quad \quad "enum":["continue","finish","satisfied","impossible","interrupt","need\_feedback"],"default":"continue"\}\},
          
    \quad \quad "defs":{"Location":\{"type":"array",
    
    \quad \quad \quad \quad \quad "description":"坐标为相对于屏幕左上角位原点的相对位置，并且按照宽高比例缩放
    
    \quad \quad \quad \quad \quad  \quad \quad  \quad \quad  \quad \quad 到0～1000，数组第一个元素为横坐标x，第二个元素为纵坐标y",
    
    \quad \quad \quad \quad \quad "items":{"type":"integer","minimum":0,"maximum":1000},"minItems":2,"maxItems":2}\}
    
\}
 
\end{CJK}

[current\_screenshot]

\tcbsubtitle{Assistant}
[thought\_and\_action]
\end{tcolorbox}

\section{System-level Usage}
\label{system-level}
\begin{figure}[!ht]
  \centering
  \includegraphics[scale=0.47]{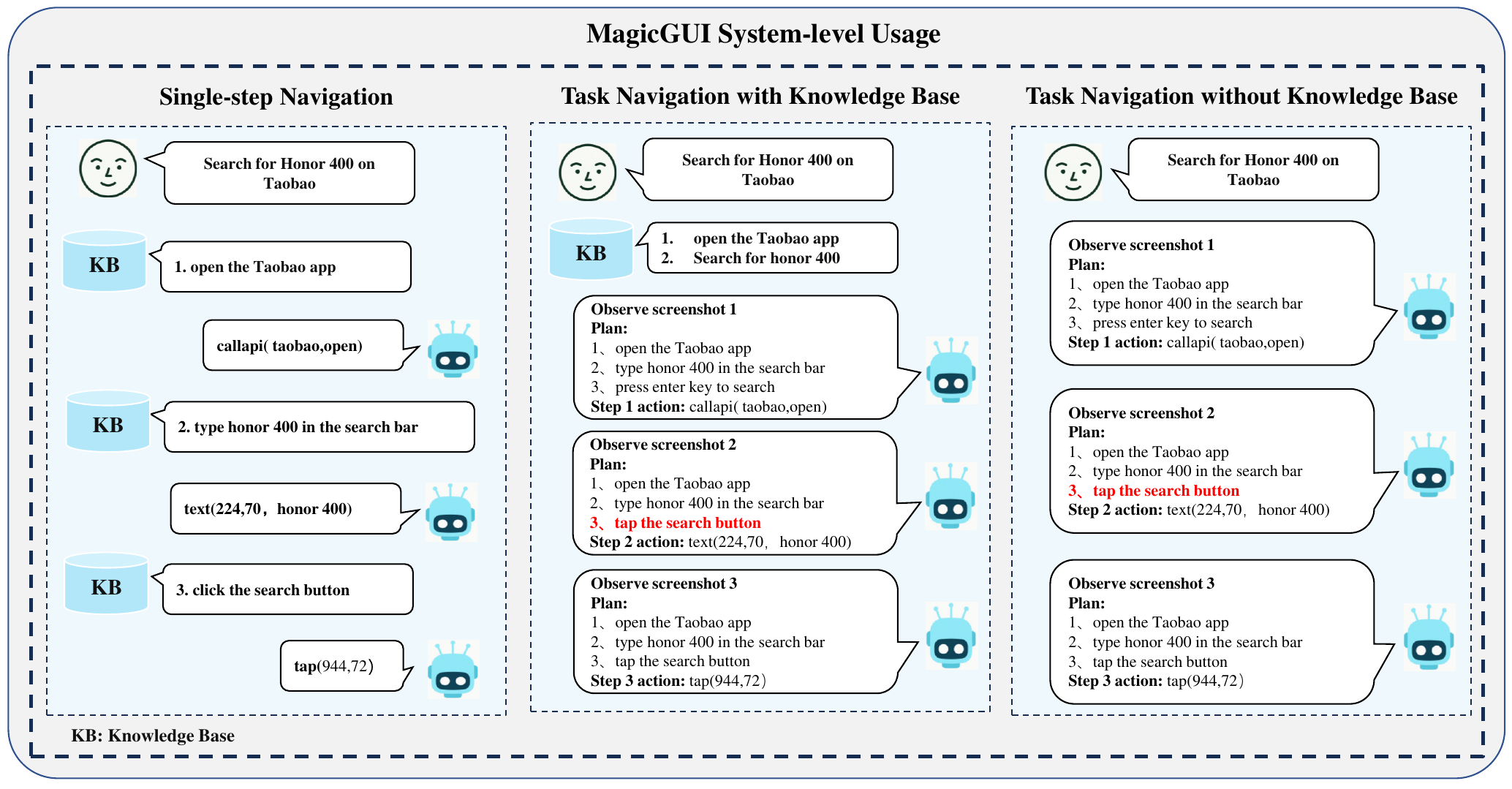}
  \caption{System-level usage case.}
  \label{fig:system-level usage}
\end{figure}

\begin{figure}[!ht]
  \centering
  \includegraphics[scale=0.49]{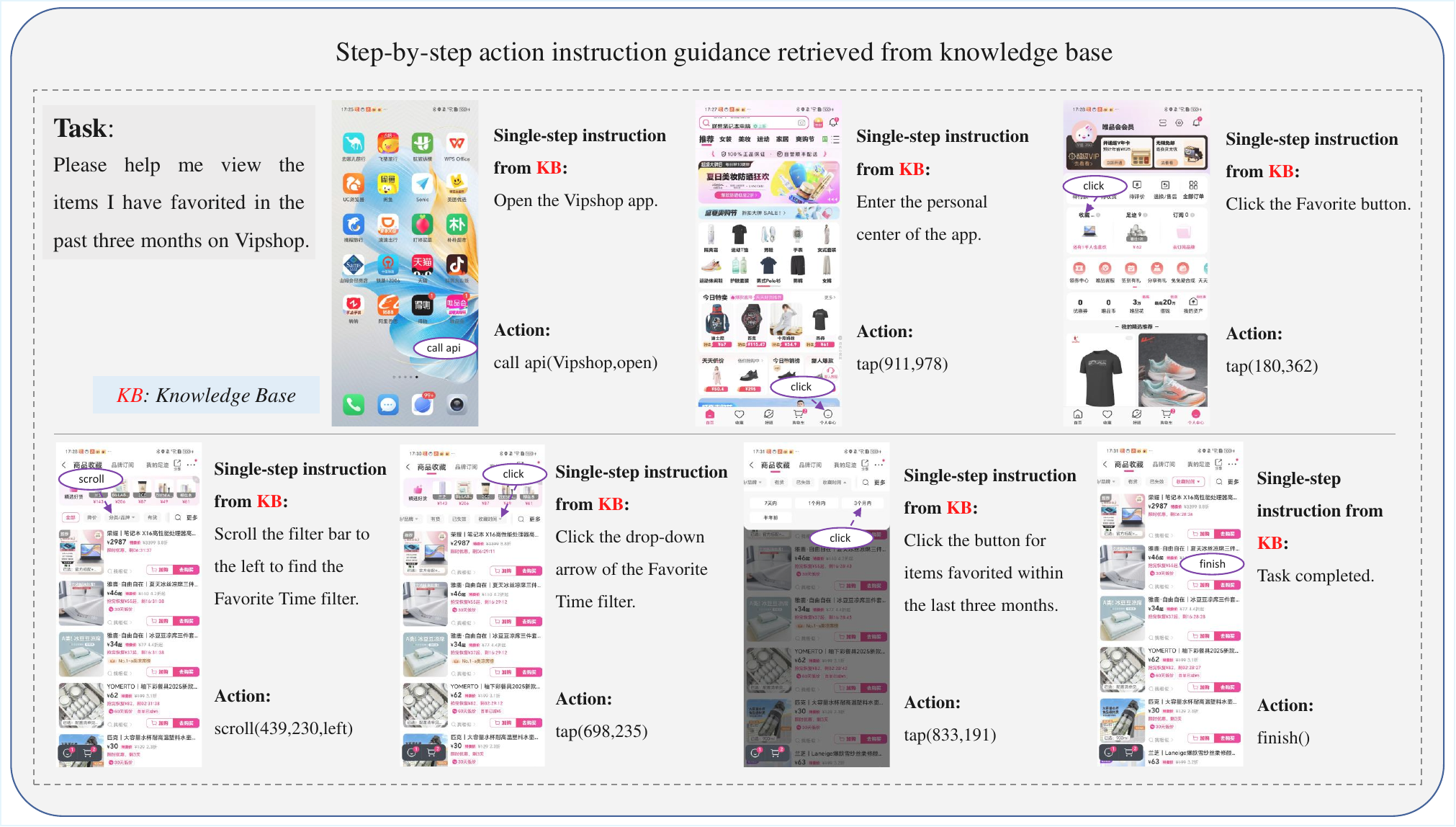}
  \caption{A demo case of single-step navigation.}
  \label{fig:case1}
\end{figure}

\begin{figure}[!ht]
  \centering
  \includegraphics[scale=0.49]{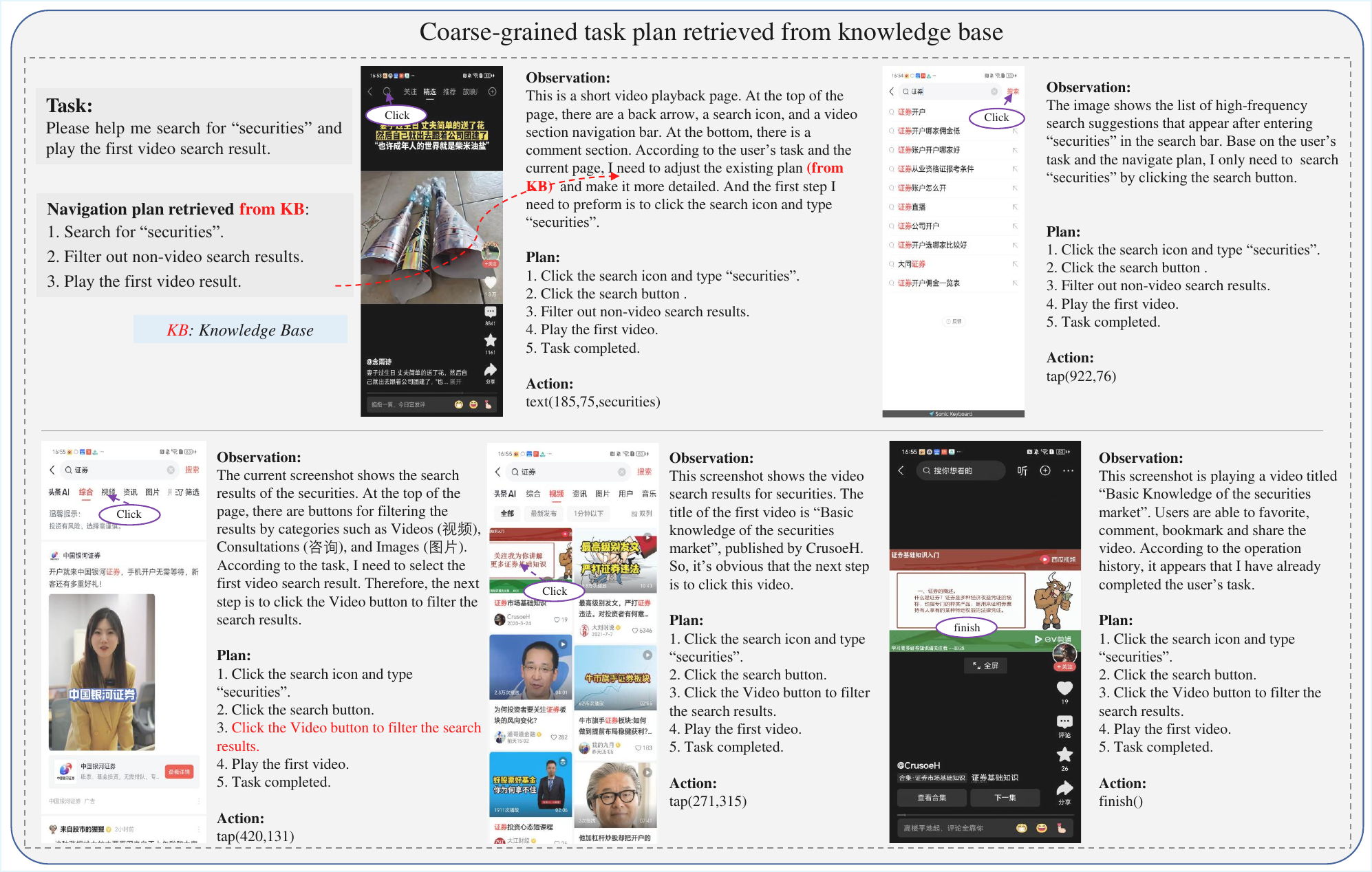}
  \caption{A demo case of task navigation with knowledge base.}
  \label{fig:case2}
\end{figure}

\begin{figure}[!ht]
  \centering
  \includegraphics[scale=0.49]{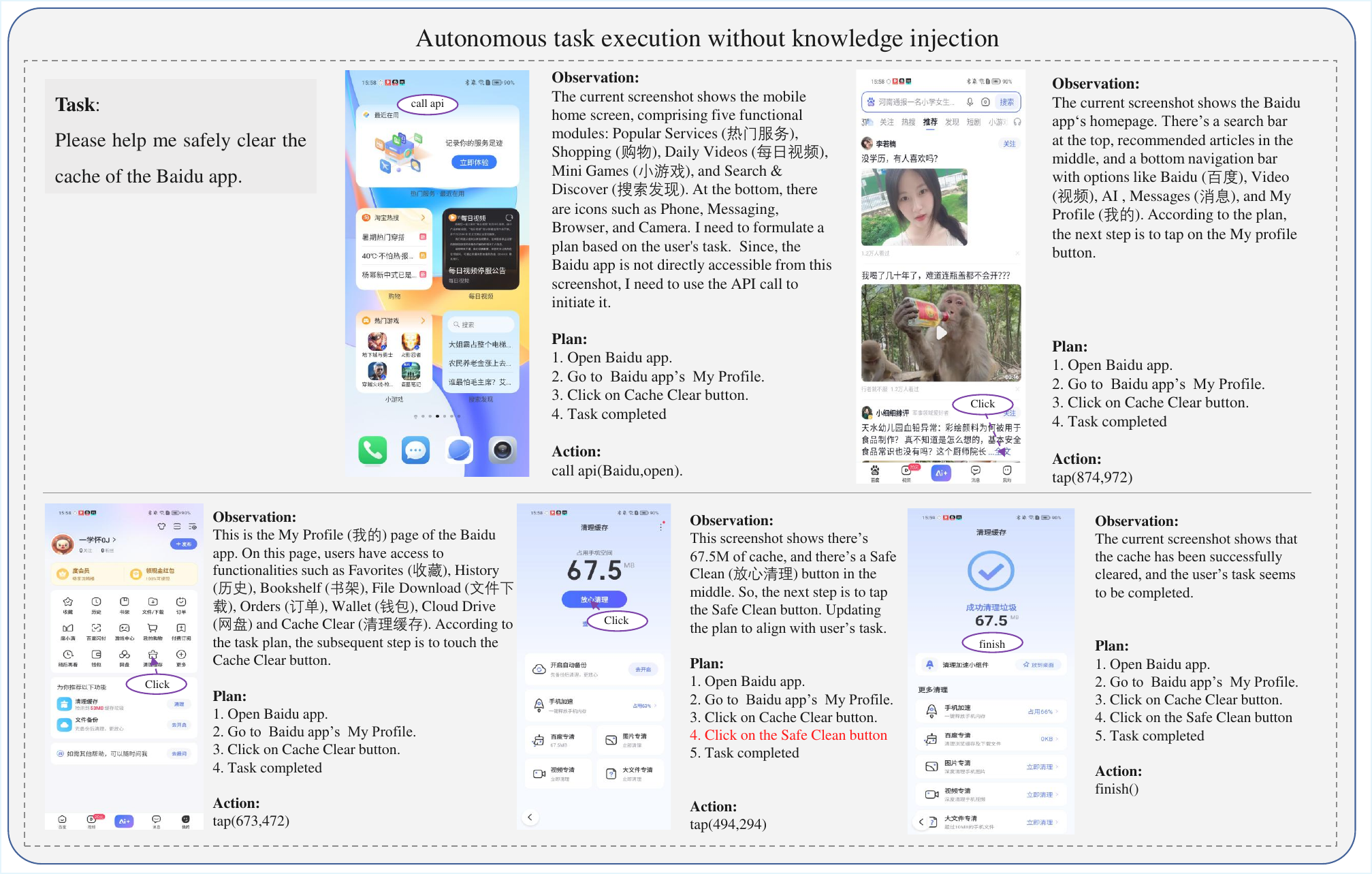}
  \caption{A demo case of task navigation without knowledge base.}
  \label{fig:case3}
\end{figure}

As shown in Figure~\ref{fig:system-level usage}, in practical business scenarios, our MagicGUI model is applied at the system level in three ways:

\begin{itemize}

\item \textbf{Single-step Navigation.} We retrieve single-step operational instructions from the knowledge base, subsequently execute these instructions using MagicGUI, and iteratively repeat this process to accomplish the user's objectives, as demonstrated in Figure~\ref{fig:case1}.

\item \textbf{Task Navigation with Knowledge Base (KB).} We retrieve a relevant navigation plan from the knowledge base based on the navigation task to guide the model in accomplishing the task. At each step of the navigation process, the model dynamically adjusts the navigation plan by considering the history of executed operations on the mobile device and the current screen state, thereby determining the precise action to be performed at that step. Through iterative refinement of the navigation plan and interaction with the device’s current screen, the model progressively completes the user’s task, as demonstrated in Figure~\ref{fig:case2}.

\item \textbf{Task Navigation without Knowledge Base (KB).} The system-level usage without a knowledge base is similar to that with a knowledge base, with the key difference being the absence of a knowledge base navigation plan for task navigation initialization, as demonstrated in Figure~\ref{fig:case3}.

\end{itemize}

\end{document}